\documentclass[
 aps,longbibliography,
 amsmath,amssymb,
 preprint,
]{revtex4-1}

\usepackage{graphicx}
\usepackage{bm}
\usepackage{natbib} 
\usepackage{amssymb,amsmath}
\usepackage{booktabs,array,dcolumn}
\usepackage{siunitx}
\usepackage{longtable}

\newcommand{\vshift}{f_{\text{shift}}}

\newcommand{\xcut}{x_{\text{cut}}}
\newcommand{\vdia}{\varphi_{\text{dia}}}
\newcommand{\vDia}{\varphi_{\text{dia}}}
\newcommand{\vsc}{\varphi_{\text{sc}}}
\newcommand{\vSc}{\varphi_{\text{sc}}}
\newcommand{\vHc}{\varphi_{\text{hc}}}
\newcommand{\vSqu}{\varphi_{\text{squ}}}

\begin{document}
\title{Dimensionality and design of isotropic interactions that stabilize honeycomb, square, simple cubic, and diamond lattices}

\author{Avni Jain}
\affiliation{McKetta Department of Chemical Engineering, University of Texas at Austin, Austin, TX 78712}

\author{Jeffrey R. Errington}
\affiliation{Department of Chemical and Biological Engineering, University of Buffalo, The State University of New York, Buffalo, New York 14260-4200}

\author{Thomas M. Truskett}
\email{truskett@che.utexas.edu}
\affiliation{McKetta Department of Chemical Engineering, The University of Texas at Austin, Austin, TX 78712}

\date{\today}

\begin{abstract} 
We use inverse methods of statistical mechanics and computer simulations to investigate whether an isotropic interaction designed to stabilize a given two-dimensional (2D) lattice will also favor an analogous three-dimensional (3D) structure, and vice versa. Specifically, we determine the 3D ordered lattices favored by isotropic potentials optimized to exhibit stable 2D honeycomb (or square) periodic structures, as well as the 2D ordered structures favored by isotropic interactions designed to stabilize 3D diamond (or simple cubic) lattices. We find a remarkable `transferability' of isotropic potentials designed to stabilize analogous morphologies in 2D and 3D, irrespective of the exact interaction form, and we discuss the basis of this cross-dimensional behavior. Our results suggest that the discovery of interactions that drive assembly into certain 3D periodic structures of interest can be assisted by less computationally intensive optimizations targeting the analogous 2D lattices.
\end{abstract}


\maketitle

Material properties are intimately linked to structural characteristics featured at various lengthscales. Thus, discovering new ways to create materials with prescribed morphologies is a key challenge in their design for specific applications. In addition to the development of top-down material fabrication strategies, there has been considerable progress in bottom-up approaches in which the primary components (molecules, nanoparticles, colloids, etc.) are engineered to promote their self-assembly into targeted structures. Examples of the latter include assembly of lithographic masks~\cite{Lithography2010Rev}, polymer membranes~\cite{ReviewDesign2012}, magnetic nanostructures~\cite{MagneticNano2006}, and colloidal superlattices~\cite{ReviewColloids2000} for photonic materials~\cite{ReviewCCPBG2010,ReviewBottomSAPC2013} to mention a few.
%

A critical part of any self-assembly design problem is understanding how tunable aspects of the interactions affect the thermodynamic stability of competing assembled states with different morphologies. For nano- to microscale particles, this understanding has been guided in part via exploratory experiments and simulations to characterize the structures that spontaneously form from systems with various particle chemistries~\cite{murray2000synthesis,QDReview2013}, shapes~\cite{glotzer2007anisotropy,ExpShapeNP2009,Glotzerscience,EscobedoNatMat2011,GeisslerNatMat2012,PtNanocubes,RevShape2013}, and surface properties~\cite{vega:9938,SurfRoughPNAS2012,ReviewpatchyMay2013,ArthiJayJPol2013}, as well as different dispersing solvents~\cite{MPPileni2011} and mixtures of assembling particles~\cite{shevchenko2006structural,royshenhar2012}. Highly-coordinated lattices with, e.g., face-centered cubic or hexagonal symmetries in three dimensions (3D)~\cite{murray2000synthesis} and triangular symmetry in two dimensions (2D)~\cite{CPC-NP-Surfaces-Interfaces}, are commonly observed in the experimental assembly of monodisperse particles with short-range, isotropic interactions. A broader array of thermodynamically stable 3D structures--including low-coordinated diamond and simple cubic lattices of interest for technological applications~\cite{DIAphoton,PCsimplecubic}--has also been demonstrated by computer simulations of monodisperse particles with softer, repulsive potentials~\cite{YKorig1,fominsc,marcottediamondpaper,edlundjcp2013,SMJainTMT2013}, including those that model the interactions between elastic spheres~\cite{hertz} or star polymers~\cite{LikosStarPot}. Similar interactions favor open 2D structures as well, including honeycomb and square lattices~\cite{EAJagla1999,PJCamp2003,GMalescio2003,JCP2dmonotonic,ZhangTorquato2013,AGP-SK-SM-2014,PRL2Dfomin2014} with, e.g., sterically-stabilized magnetic particles in the presence of an external field~\cite{PZiherlPRL2007} providing one novel experimental realization. Finally, low-coordinated lattices can also be stabilized by particles with patchy surfaces or faceted shapes, as demonstrated by experiments (mostly in 2D~\cite{SteveGNature,WHEvers2012}) and simulations (in both 2D~\cite{2DPatchydesign,ATselfassembly2014} and 3D~\cite{vega:9938,PCCPPatchy-perspective,sciortinonatcom,ReviewpatchyMay2013}). For a given application, the choice of self-assembling components often hinges on practical considerations including the complexity and expense associated with particle synthesis and the kinetics of assembly.  
%

Despite the fact that various interaction models are known to stabilize specific lattices of interest in a given spatial dimension (2D or 3D), much less is understood about how spatial dimension affects the design rules for assembly. For example, to what extent will an interaction designed to stabilize a given 2D lattice also favor an analogous 3D structure, and vice versa~\footnote{ The term `analogous structures' refers to the pairs of 2D-3D lattices (e.g., honeycomb-diamond and square-simple cubic) that have specific coordination-shell similarities that can allow a single isotropic pair potential to favor the stability of both. This structural similarity is addressed both in the discussion of Fig.~3 and Fig.~S1 (in the Supplementary Information)}? The answer is of fundamental interest and may also have important practical implications because finding interactions that stabilize lattices in 2D is a simpler and less computationally demanding material design problem than in 3D. Here, we study this question using computer simulations and model potentials designed by inverse statistical mechanical optimization~\cite{TorquatoRev,AIChEPersp}.
%

In particular, we determine the 3D ordered lattices favored by models with isotropic potentials~\(\vHc\) (or \(\vSqu\)) optimized to exhibit stable 2D honeycomb (or square) periodic structures, as well as the 2D ordered structures favored by isotropic interactions~\(\vdia\) (or \(\vSc\)) designed to stabilize 3D diamond (or simple cubic) lattices~\footnote{We note that particles confined to a 2D monolayer, such as at a liquid-liquid interface or on a substrate, may interact via an effective pair potential that is different from the one that the same particles experience in a 3D bulk fluid.}. As we show, the isotropic potentials optimized for either 2D or 3D target structures also do surprisingly well at stabilizing the analogous lattices in the other dimension.  
%

A specified target lattice is the ground state for a given pair potential \(\varphi\) and pressure \(p\) if, and only if, it is mechanically stable at this condition and its zero-temperature chemical potential (i.e., molar enthalpy) is lower than that of all other mechanically stable competing structures. Here, we use a stochastic optimization approach (described in detail elsewhere~\cite{SMJainTMT2013}) to discover new model pairwise interactions \(\varphi_{\text{target}}\) that maximize the range of density \(\rho\) for which a 2D target lattice is the ground state. In our optimizations, we consider isotropic, convex-repulsive pair potentials that qualitatively mimic the soft, effective interactions of sterically-stabilized colloids or nanoparticles~\cite{schapotschnikow2008molecular}. The form we adopt can be expressed~\cite{SMJainTMT2013}
\begin{equation}
\begin{split}
\varphi(x) &= \epsilon \{ A x^{-n}+\sum_{j=1}^2\lambda_{j}\left(1-\tanh\left[k_{j}\left(x-\delta_j\right)\right]\right) + \vshift (x)\} H[\xcut-x]  
\end{split}
\label{eq:potential}
\end{equation}
Here, \(x=r/\sigma\) is a dimensionless interparticle separation; \(\epsilon\) and \(\sigma\) are characteristic energy and length scales; \(\xcut\) is the dimensionless potential range; \(H\) is the Heaviside step function; and \(\vshift (x)=Px^2+Qx+R\) is a shifting function with fitting constants {\(P, Q, R\)} chosen to ensure \(\varphi(\xcut)=\varphi^{\prime}(\xcut)=\varphi^{\prime \prime}(\xcut)=0\). All together, there are nine dimensionless parameters that can be varied in the optimization algorithm (\(\xcut, A, n, \lambda_{1},k_{1},\delta_{1},\lambda_{2},k_{2},\delta_{2}\)); however, one is not independent of the others because we also require \(\varphi(1)/\epsilon=1\). From here forward, we report quantities implicitly nondimensionalized by appropriate combinations of \(\epsilon\) and \(\sigma\). 
%

To identify the ground-state phase diagram for a given pair potential \(\varphi\), we compare the \(p\)-dependent, zero-temperature chemical potentials of a wide variety of Bravais and non-Bravais lattices in a `forward' calculation. Several methods for identifying candidate ground states are available, including evolutionary optimization~\cite{GALikos1,bianchiJCP12} as well as shape matching and machine learning algorithms~\cite{GregVoth2013}. In this study, we use simulated annealing optimization~\cite{SMJainTMT2013} to determine free lattice parameters which minimize the chemical potentials of the structures subject to the constraint of mechanical stability, as determined by phonon spectra analysis~\cite{ashcroft1976solid}. In 2D, the Bravais lattices consist of oblique, rhombic, square, rectangular, and triangular symmetries; here, we limit our consideration of non-Bravais lattices to honeycomb, kagome, and other five-vertex semi-regular tilings, namely snub-hexagonal, snub-square, and elongated-triangular. For 3D, we consider the following Bravais and non-Bravais lattices identified in a previous study on closely related model interactions~\cite{zerotempsanti}:  face-centred cubic (FCC), body-centred cubic (BCC), simple cubic (SC), diamond, pyrochlore, body-centred orthogonal (BCO), hexagonal (H), rhombohedral (hR), cI16, oC8, \(\beta\)Sn, A7, A20, and B10. While the methods employed both to determine the interaction potentials optimal for a target lattice and to compute the corresponding ground states are identical in 2D and 3D, we note that calculations are significantly faster in 2D than in 3D due to the smaller number of competing structures to consider in 2D and the reduced dimensionality of the lattice sum and the phonon spectra evaluations.
%

For computational efficiency of inverse optimizations in 2D or 3D, only a limited set of competing structures can be considered for a specific target lattice, ideally consisting of the lattices which have the lowest chemical potentials for the interaction type over the density range of interest. Here, we use a simple iterative process for determining the competitive lattice pools. Specifically, we 
(1) begin with a trial set of competitive structures; 
(2) carry out an inverse optimization calculation using this competitive pool to obtain parameters for a trial optimal potential; 
(3) perform an extensive forward calculation to determine the ground-state phase diagram of the trial potential; 
(4) as necessary, refine the competitive pool based on the lattices that appear in the forward calculation in (3) and return to step (2). The final pools determined from this method contained a diverse array of structures in 2D and 3D~\footnote{Using the approach outlined in the text, the competitive pools determined for use in optimizations targeting the honeycomb lattice consisted of triangular, square, kagome, snub-square, elongated triangular, rectangular (\(b/a=1.49\)), rectangular (\(b/a=1.54\)), rectangular (\(b/a=1.56\)), rectangular (\(b/a=1.7\)), and snub-hexagonal lattices. For the square lattice, the final pool comprised triangular, oblique (\(b/a=1.514\), \(\theta\)=1.234), kagome, honeycomb, elongated triangular, snub-hexagonal, and snub-square lattices. For diamond, the pool~\cite{SMJainTMT2013} consisted of FCC, WUR, SH \((c/a=1.5)\), \(\beta\)Sn \((c/a=1.39)\), \(\beta\)Sn \((c/a=1.25)\), A7 \((b/a=3.79, u=0.1385)\), and A20 \((b/a=1.728, c/a=0.626,y=0.167)\) lattices. For simple cubic, the pool~\cite{SMJainTMT2013} comprised FCC, BCC, DIA, SH \((c/a=1)\), SH \((c/a=1.08)\), SH \((c/a=1.172)\), A20 \((b/a=1.72, c/a= 0.66, y= 0.67)\), \(\beta\)Sn \((c/a=0.873)\), \(\beta\)Sn \((c/a=0.78)\), and \(\beta\)Sn \((c/a=1.75)\) lattices. Here, \(b/a\) and \(c/a\) denote the aspect ratio of the sides of the unit cell, and \(\theta\) is the angle between the two sides. The other symbols \(u\) and \(y\), we adopt here, are the same as those used in a previous study~\cite{zerotempsanti}.}.
%

To obtain information about the thermal stability of the target lattices, we also perform Monte Carlo quench simulations in which a high-temperature fluid is instantaneously cooled down to a much lower temperature to observe assembly of the target structure.
Our simulation sizes were chosen such that larger systems did not affect the results (for more details, see Table S1 and discussion in Supplementary Information). We note that interactions previously optimized to stabilize 3D target ground states of diamond (\(\vdia\)) and simple cubic (\(\vsc\)) lattices over a wide range of density--using methods identical to those employed here--lead to target crystalline phases with good thermal stability~\cite{jainJCP}.
%
    
The interaction potentials we obtain for maximizing the density range of 2D honeycomb- and square-lattice ground states~\footnote{Optimal parameters for the honeycomb-forming interaction \(\vHc\) are \(A=0.326914\), \(n=3.63306\), \(\lambda_{1}=0.286436\), \(k_{1}=3.6569\), \(\delta_{1}=1.26977\), \(\lambda_{2}=1.03258\), \(k_{2}=3.03683\), \(\delta_{2}=1.1557\), \(P=-0.175071\), \(Q=0.800825\), \(R=-0.937142\), and \(\xcut=2.03291\). 
For the square-forming potential \(\vSqu\), they are \(A=0.0946889\), \(n=3.53953\), \(\lambda_{1}=0.32989\), \(k_{1}=1.89197\), \(\delta_{1}=1.99003\), \(\lambda_{2}=0.062012\), \(k_{2}=5.89983\), \(\delta_{2}=1.0809\), \(P=0.433908\), \(Q=-2.4391\), \(R=3.46552\), and \(\xcut=2.27813\).} together with previously optimized interactions for diamond- and simple cubic-lattice ground states~\cite{SMJainTMT2013}, are shown in Fig.~1. Notice that interactions \(\vHc\) and \(\vdia\) are remarkably similar to one another, despite the fact that they were obtained from optimizations favoring different (albeit analogous) structures in different spatial dimensions. 
As is shown in the inset to Fig.~1, significant discrepancies between these potentials (i.e., the steeper repulsions of \(\vHc\)) are only present for interparticle separations \(x<0.6\) that, as we confirm below, are closer than the nearest neighbor distance for the honeycomb or diamond lattices in the density range where the structures are stable for either model. Based on the similarity of these interactions, one might already expect that \(\vHc\) and \(\vdia\) would stabilize similar lattices in 2D and 3D. On the other hand, we see appreciable differences between the potentials \(\vSqu\) and  \(\vsc\) optimized to stabilize 2D square and 3D simple cubic lattices, respectively. Of the four interactions studied here, \(\vSqu\) has the softest repulsive core and the longest range, while \(\vsc\) has the steepest core repulsion and the shortest range.
%

In Fig.~2, we show the results of our forward calculations, i.e., the 2D ground states for the four optimized potentials as a function of density~\footnote{Also see Table S2 in Supplemental Information which tabulates the stable ground-state lattices with their corresponding density ranges and lattice parameters}. Shaded regions represent densities where the ground state comprises two neighboring lattices in coexistence. First, we note that the 2D inverse optimization calculations succeed in their goal: stable honeycomb- and square-lattice ground states appear for \(\vHc\) and \(\vSqu\), respectively, over very wide density ranges, especially when compared to those of other repulsive, isotropic interaction models~\cite{JCP2dmonotonic,2DMonotonicSM,PRL2Dfomin2014} known to form these phases. Perhaps more noticeable is not only that the 2D honeycomb lattice is stabilized over a similar density range by the 3D-optimized \(\vdia\) (a result now expected based on the similarity to \(\vHc\) shown in Fig.~1), but also that the square lattice is stabilized over a wide density range by \(\vsc\) (despite significant differences compared to \(\vSqu\)). In other words, for both cases, stable 2D ground states of interest were obtainable by optimizing interactions for a corresponding analogous target lattice in 3D.
%

To test the same approach in the other direction, i.e., whether optimizing analogous 2D structures will stabilize 3D target lattices of interest, we also determine the 3D ground states for \(\vHc\) and \(\vSqu\). The results, presented in Table~I, show that \(\vHc\) and \(\vSqu\) indeed display wide stability regions for diamond and simple cubic lattice ground states, respectively. In fact, not only are the density ranges of the stable diamond lattice comparable for \(\vHc\) and \(\vdia\), but the density range of the simple cubic lattice for \(\vSqu\) is even slightly wider than that of \(\vSc\)~\footnote{Ground states and finite-temperature phase boundaries for \(\vdia\) and \(\vsc\) have been determined previously and are presented in detail elsewhere~\cite{SMJainTMT2013,jainJCP}.}. The latter result likely reflects the fact that the faster optimizations targeting 2D ground states enables a more thorough exploration of parameter space during the calculation than is practical in the 3D optimizations. 
%

That particles with isotropic interactions encoded to form 3D diamond (or simple cubic) lattices also display 2D honeycomb (or square) arrays, although nontrivial, is in some sense not surprising. The tetrahedrally-coordinated diamond lattice itself consists of undulating interconnected trivalent honeycomb networks, and the simple cubic structure comprises square arrays stacked in registry. However, the outcome that particles with interactions designed to stabilize 2D honeycomb (or square) lattices also favor diamond (or simple cubic) lattices {\emph{and not other morphologies containing honeycomb (or square) motifs such as graphite (or body-centered cubic) structures}} is much more interesting.
%

To understand these results, it is helpful to recall that--for isotropic potentials--the zero-temperature chemical potential depends only on the pair interaction and properties of coordination shells located at distances closer than the interaction cut-off, \(x<\xcut\). In Fig.~3, we plot the interparticle separations corresponding to the first, second, and third coordination shells \{\(x_{1},x_{2},x_{3}\)\} for the four lattices of interest here--honeycomb (hc), square (squ), diamond (dia) and simple cubic (sc)--considering densities where these lattices are the ground states for the models \(\vHc\) and \(\vSqu\). First, note that there is considerable overlap between the coordination-shell distances of the honeycomb and diamond structures. Thus, an isotropic potential which stabilizes a honeycomb structure in 2D is expected to be an excellent (if not necessarily optimal) candidate for forming a diamond lattice in 3D, and vice versa. This helps to explain the near identical potentials, \(\vHc\) and \(\vDia\), despite their being obtained via optimization of different target structures in different spatial dimensions.
%

To gain further insights, we also compare the coordination-shell distances of the honeycomb lattice with another related 3D structure, graphite, which consists of stacks of 2D honeycomb (i.e., graphene) sheets. Note that only the nearest-neighbor distances of mechanically stable 3D graphite lattices align with the first coordination-shell separations of 2D honeycomb structures, and there is substantial mismatch of other relevant coordination distances (\(i\geq2\)) (see Fig.~S1 in the Supplementary Information). In this important sense, graphite--while closely related to the honeycomb lattice in other ways--is not as {\emph{analogous}} to honeycomb as the 3D diamond structure is in its relation between interaction and coordination-shell structure, and is thus, not favored as a ground-state by \(\vHc\) at any density. In comparing the other case of square versus simple cubic lattices, we see that the first two coordination shells of these structures similarly overlap, but the third shell positions are not in alignment. This result--together with the ground-state calculations presented above--suggests that, for short-range interactions, the common separation distances between the nearest and next-nearest neighbors for square and simple cubic structures in enough to allow for an optimal 2D square-forming potential to assemble into 3D simple cubic structures, and vice versa. However, the differences in the third-shell distances might help to explain the significant variations in the optimized potentials targeting 2D-square (\(\vSqu\)) versus 3D-simple cubic (\(\vSc\)) lattices shown in Fig.~1.
%

In Fig~4a-d, we present snapshots of configurations obtained from the Monte Carlo quench simulations for the four potential models. Configurations for the 3D diamond and simple cubic lattice obtained via quenching systems interacting with 2D-optimized \(\vHc\) and \(\vSqu\) interactions are shown in Fig~4e and Fig~4f, respectively. 
The structures obtained were inspected visually, and their configurational energies and pair distribution functions \(g(r)\) were compared to equilibrated lattice structures at the corresponding densities and temperatures (see Table S1). 
In Fig.~4g, the complete overlap of the pair distribution functions of the quenched fluid (red circles) and the equilibrated simple cubic structure (black dashed lines) demonstrates the assembly of a defect-free simple cubic crystal. The \(\vHc\) model similarly assembles into a (slightly defective) diamond structure as illustrated by the comparison of the pair distribution functions in Fig.~4h. The energy of the quenched configuration is only \(~0.09\%\) higher than the perfectly equilibrated diamond lattice. Nonetheless, in all cases, the structures obtained by the Monte Carlo quench procedure match the expectations of the ground-state calculations.
%

To summarize, we have investigated the cross-dimensional phase behavior of specifically designed isotropic interactions with low coordination. In particular, we have determined the 3D ordered lattices favored by isotropic potentials \(\vHc\) (or \(\vSqu\)) optimized to exhibit stable 2D honeycomb (or square) lattice structures, as well as the 2D periodic structures favored by isotropic potentials \(\vdia\) (or \(\vSc\)) optimized to assemble into 3D diamond (or simple cubic) morphologies. We find surprising transferability of interactions designed to stabilize analogous structures in 2D and 3D, and we gain insights into this behavior by studying the different ways in which information in the analogous target structures encodes itself in the optimal isotropic potentials through the coordination-shell geometry. 
%

One practical implication of the observed physics in this study is that the design of certain 3D lattices can greatly benefit from knowledge of potentials derived to maximize the stability of analogous 2D structures, information which can be obtained at relatively modest computational expense. The computational efficiency gained from this approach might be most valuable in multi-step optimization processes, where the goal to search for an interaction potential favoring a target structure is only one of several objectives within the design calculation. It will also be interesting in future studies to explore the effects of the interaction range on the cross-dimensional behavior of isotropic interactions obtained through inverse design, especially where one limits the potential range to encompass only two coordination shells. While we focus here on the dimensionality dependence of design rules pertaining to target structures formed by isotropic interactions, it will also be informative to study the effect of spatial dimension on other classes of interactions, e.g., short-ranged anisotropic interactions of patchy particles relevant to 2D and 3D assembly scenarios. 
%

Finally, in the context of cross-dimensional freezing behavior, we note the differences between the soft repulsive interactions studied here--which enthalpically stabilize low-coordinated periodic structures--and hard-sphere systems where entropy drives the particles to adopt close-packed periodic structures at high density. For the latter, crystallization from the fluid becomes increasingly more challenging in higher spatial dimensions due to correspondingly stronger geometric frustration~\cite{PRE2009CharbFrenk1,PRE2009CharbFrenk2}. The role that frustration plays in the dimensionality dependence of crystallization for particles with considerably softer repulsions remains a potentially rich area for future study.  

T.M.T. acknowledges support of the Welch Foundation (F-1696) and the National Science Foundation (CBET-1403768). J.R.E. acknowledges support of the National Science Foundation (CHE-1012356).  We also acknowledge the Texas Advanced Computing Center (TACC) at The University of Texas at Austin for providing HPC resources that have contributed to the research results reported within this paper.


\begin{center}
\begin{longtable}{lll}
\caption{\label{tab:PD-DIA-HCopt-pot}3D ground states for \(\vHc\) and \(\vSqu\) with their corresponding density ranges and optimal lattice parameters. Roman numerals denote different structures of the same lattice type. Nomenclature is that of an earlier reference~\cite{zerotempsanti}.} \\
\hline 
\endfirsthead
Lattice & Stability range & Lattice parameters \\
\hline
\endhead
  \multicolumn{3}{l}{\emph{Honeycomb-lattice forming potential}, \(\vHc\)} \\
   \hline
   \noalign{\smallskip}
   BCC         & \([0.589,0.677]\)       \\
   A7-I        & \([0.725,0.777]\)      & \(b/a:\>2.23\), \(u:\>0.075\) \\
   A7-II       & \([0.856,1.05]\)       & \(b/a:\>4\), \(u:\>[0.68,0.81]\) \\
   \textbf{Diamond} & \textbf{[1.091,1.376]}  \\
   Hexagonal   & \([1.468,1.474]\)      & \(c/a:1.38\) \\
   A20-I       & \([1.477,1.498]\)      & \(b/a:\>1.72\), \(c/a:\>2.8\),\(\,y:\>0.5\) \\ 
   A20-II      & \([1.54,1.851]\)       & \(b/a:\>1.8\), \(c/a:\>0.65\),\(\,y:\>0.66\) \\ 
   \noalign{\smallskip}
   \hline
   \multicolumn{3}{l}{\emph{Square-lattice forming potential}, \(\vSqu\)} \\
   \hline
   \noalign{\smallskip}
   \(\beta\)Sn-I  & \([0.587,0.641]\) & \(c/a:2.67\)             \\
   B10-I          & \([0.664,0.798]\) & \(c/a:\>0.4\),\(z:\>0.5\)\\
   FCC            & \([0.828,0.961]\)                           \\
   A20-I          & \([0.991,1.047]\) & \(b/a:\>1.0\), \(c/a:\>0.68\),\(\,y:\>0.8\) \\ 
   oC8-Ga         & \([1.056,1.094]\) & \(b/a:\>1.0\), \(c/a:\>1.5\) \\
                  &                   & \(\,u:\>0.75\),\(\,v:\>0.163\) \\
   B10-II         & \([1.1,1.267]\)   & \(c/a:\>[0.72,0.73]\),\\
                  &                   & \(z:\>[0.38,0.39]\) \\
   A20-II         & \([1.282,1.298]\) & \(b/a:\>2.29\), \(c/a:\>1.79\),\(\,y:\>0.08\) \\ 
   \(\beta\)Sn-II & \([1.322,1.478]\) & \(c/a:\>[\,0.57,0.64]\) \\
   Hexagonal-I    & \([1.49,1.592]\)  & \(c/a:\>[\,0.887,0.9]\) \\
   \textbf{Simple cubic} & \textbf{[1.606,1.949]}                 \\
   \(\beta\)Sn-III  & \([1.98,2.106]\)  & \(c/a:2.74\)            \\
   \hline
  \end{longtable}
\end{center}

\begin{figure}
  \includegraphics{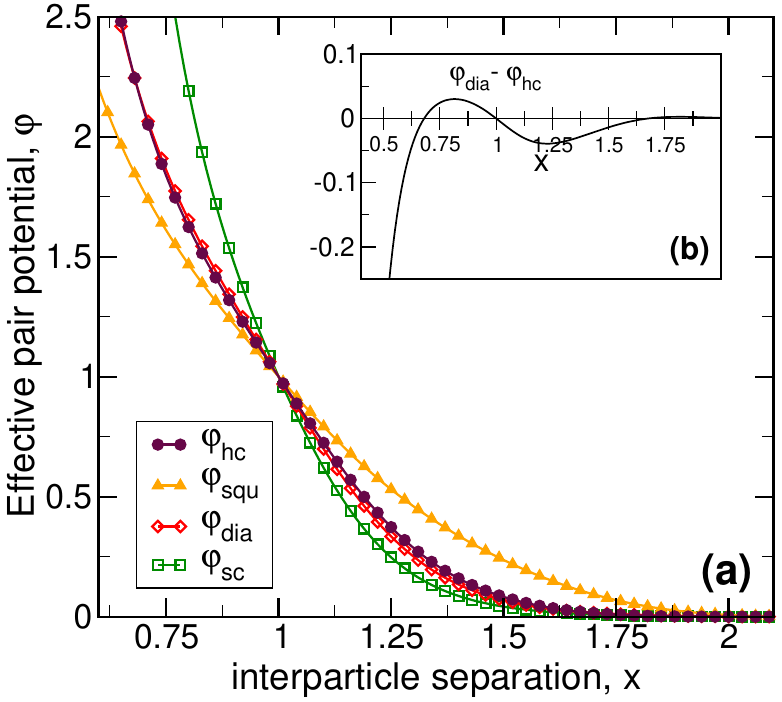}
  \caption{Isotropic, convex-repulsive potentials, \(\vHc\) and \(\vSqu\) (described in the text), which maximize the density range of mechanically stable 2D honeycomb- and square-lattice ground states, respectively. Also shown are previously designed potentials, \(\vdia\) and \(\vsc\),\cite{SMJainTMT2013} that maximize the density range of mechanically stable 3D diamond- and simple cubic-lattice  ground states, respectively. The inset highlights subtle differences between \(\vdia\) and \(\vHc\).}
  \label{fig:Potentials}
    \end{figure}

\begin{figure*}
  \includegraphics{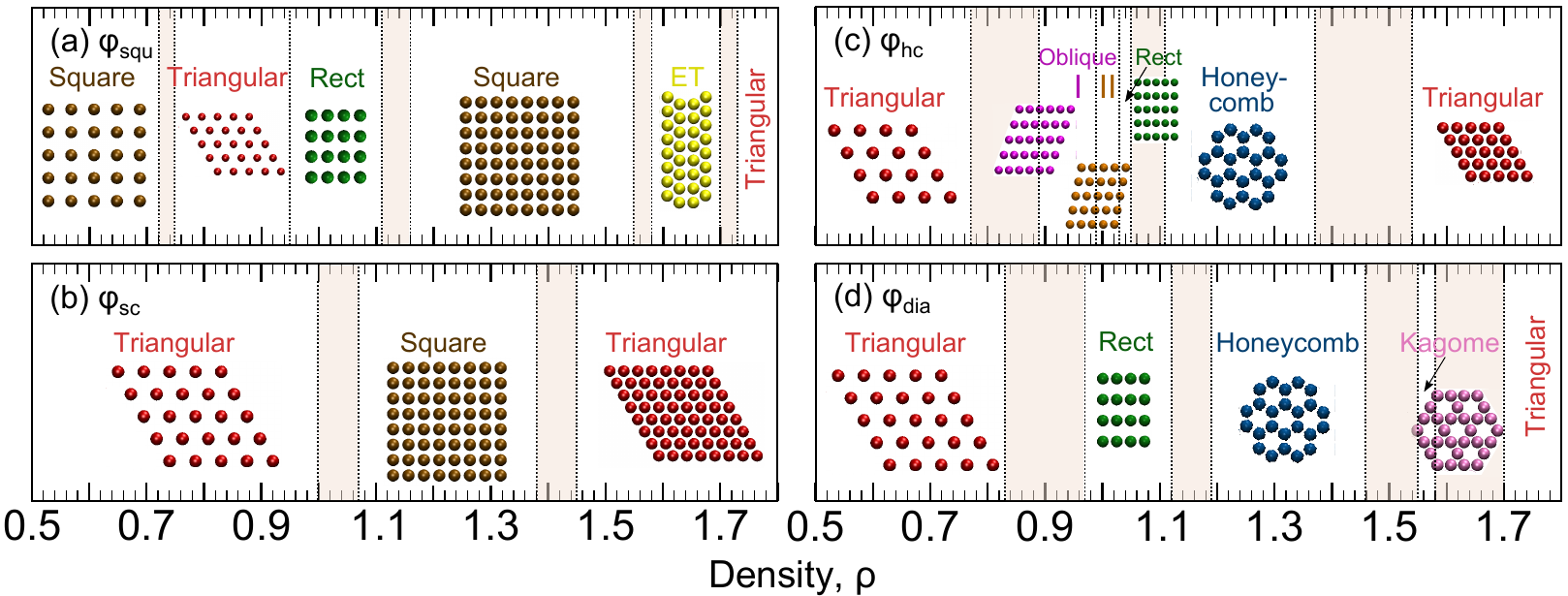}
  \caption{
  2D ground-state lattices as a function of density for (a) \(\vSqu\), (b) \(\vsc\), (c) \(\vHc\), and (d) \(\vdia\). Shaded regions represent coexistence between the neighboring lattices on the phase diagram. ET represents the elongated triangle Archimedean tiling. Parameters for the oblique and rectangular (Rect) lattices are provided in Table S2 in Supplementary Information.}
  \label{fig:2DPhases}
  \end{figure*}

\begin{figure}
 \includegraphics{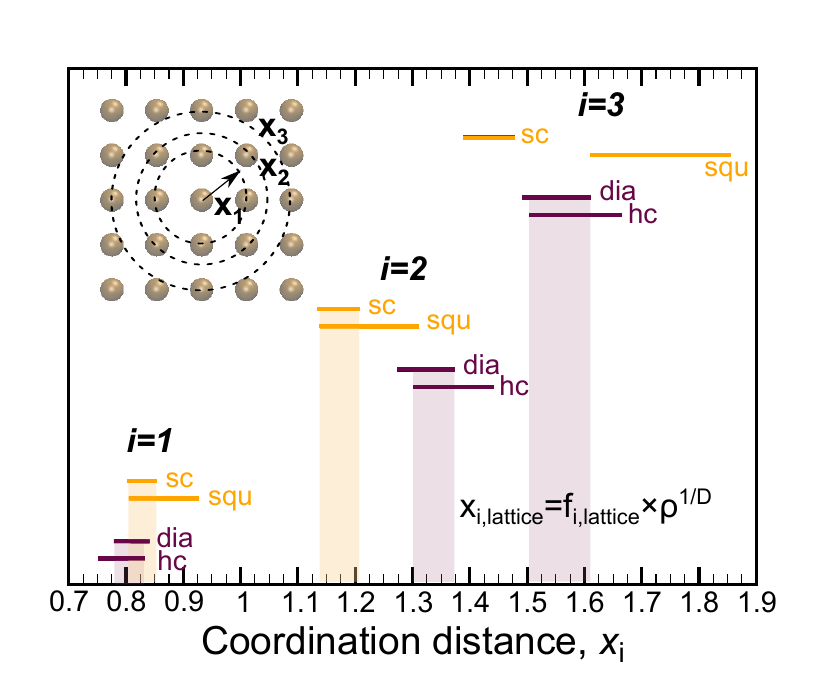}
  \caption{Separations corresponding to first (\(i=1\)), second (\(i=2\)), and third (\(i=3\)) coordination shells corresponding for stable densities of honeycomb (hc) \(\rho=[1.11,1.37]\) and diamond (dia) \(\rho=[1.09,1.38]\) ground states for \(\vHc\), as well as square (squ) \(\rho=[1.16,1.55]\) and simple cubic (sc) \(\rho=[1.61,1.95]\) ground states for \(\vSqu\). 
The coordination distance (\(x_i\)) for shell \(i\) is related to density (\(\rho\)) by 
\(x_{\text{i,lattice}}=f_{\text{i,lattice}}\times\rho^{\text{1/D}}\), where \(D=2\) for hc and squ, and \(D=3\) for dia and sc. 
For hc, \(f_{1}=0.877383, f_{2}=\sqrt{3}f_{1}, f_{3}=2f_{1}\); 
for squ, \(f_{1}=1, f_{2}=\sqrt{2}f_{1}, f_{3}=2f_{1}\);
for dia, \(f_{1}=0.866025, f_{2}=\sqrt{8/3}f_{1}, f_{3}=\sqrt{11/3}f_{1}\);
for sc, \(f_{1}=1, f_{2}=\sqrt{2}f_{1}, f_{3}=\sqrt{3}f_{1}\). Shaded regions indicate overlap of coordination shell distances for analogous 2D and 3D stable lattice ground states.}
  \label{fig:Coordshells}
\end{figure}

\begin{figure}
  \includegraphics{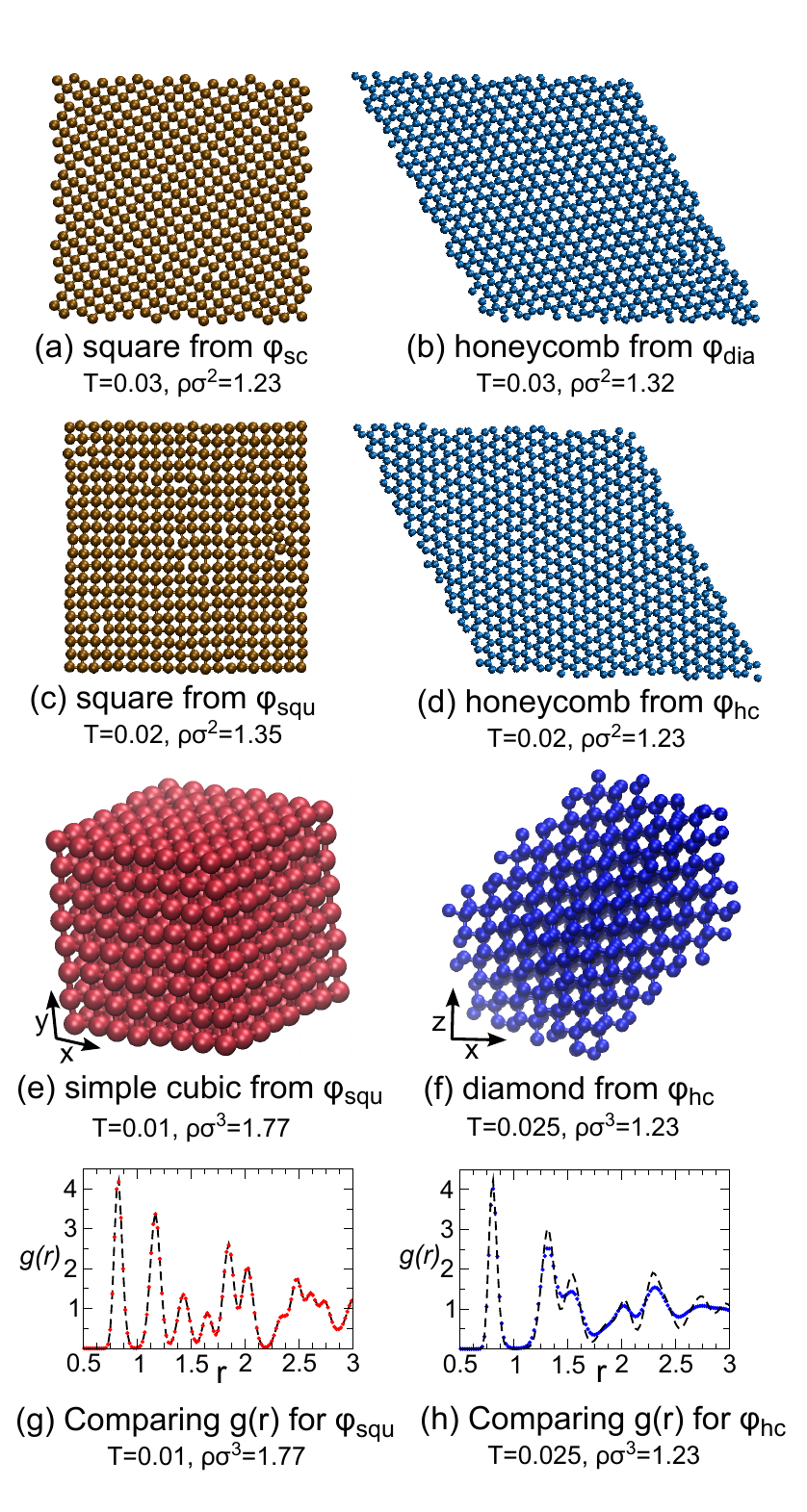}
  \caption{Snapshots of 2D configurations obtained from Monte Carlo quenches of (a) \(\vSc\), (b) \(\vDia\), (c) \(\vSqu\), (d) \(\vHc\), and 3D configurations obtained from quenches of (e) \(\vSqu\), and (f) \(\vHc\). For the 3D structures, we plot the corresponding pair distribution functions g(r) of the quenched configurations (dots) and the perfectly equilibrated lattice structures (dashed lines) in (g) and (h), at the specific temperature (T) and density (\(\rho\)), to demonstrate the structural similarities. The similarity between the annealed and perfectly equilibrated diamond lattice in (h) shows that the annealed structure is not quite defect-free but the configuration energy is only 0.09\% higher than that of the perfect lattice. See Table S1 in Supplementary Information.}
  \label{fig:Snapshots}
\end{figure}


\begin{thebibliography}{64}%
\makeatletter
\providecommand \@ifxundefined [1]{%
 \@ifx{#1\undefined}
}%
\providecommand \@ifnum [1]{%
 \ifnum #1\expandafter \@firstoftwo
 \else \expandafter \@secondoftwo
 \fi
}%
\providecommand \@ifx [1]{%
 \ifx #1\expandafter \@firstoftwo
 \else \expandafter \@secondoftwo
 \fi
}%
\providecommand \natexlab [1]{#1}%
\providecommand \enquote  [1]{``#1''}%
\providecommand \bibnamefont  [1]{#1}%
\providecommand \bibfnamefont [1]{#1}%
\providecommand \citenamefont [1]{#1}%
\providecommand \href@noop [0]{\@secondoftwo}%
\providecommand \href [0]{\begingroup \@sanitize@url \@href}%
\providecommand \@href[1]{\@@startlink{#1}\@@href}%
\providecommand \@@href[1]{\endgroup#1\@@endlink}%
\providecommand \@sanitize@url [0]{\catcode `\\12\catcode `\$12\catcode
  `\&12\catcode `\#12\catcode `\^12\catcode `\_12\catcode `\%12\relax}%
\providecommand \@@startlink[1]{}%
\providecommand \@@endlink[0]{}%
\providecommand \url  [0]{\begingroup\@sanitize@url \@url }%
\providecommand \@url [1]{\endgroup\@href {#1}{\urlprefix }}%
\providecommand \urlprefix  [0]{URL }%
\providecommand \Eprint [0]{\href }%
\providecommand \doibase [0]{http://dx.doi.org/}%
\providecommand \selectlanguage [0]{\@gobble}%
\providecommand \bibinfo  [0]{\@secondoftwo}%
\providecommand \bibfield  [0]{\@secondoftwo}%
\providecommand \translation [1]{[#1]}%
\providecommand \BibitemOpen [0]{}%
\providecommand \bibitemStop [0]{}%
\providecommand \bibitemNoStop [0]{.\EOS\space}%
\providecommand \EOS [0]{\spacefactor3000\relax}%
\providecommand \BibitemShut  [1]{\csname bibitem#1\endcsname}%
\let\auto@bib@innerbib\@empty
\bibitem [{\citenamefont {Zhang}\ \emph {et~al.}(2010)\citenamefont {Zhang},
  \citenamefont {Li}, \citenamefont {Zhang},\ and\ \citenamefont
  {Yang}}]{Lithography2010Rev}%
  \BibitemOpen
  \bibfield  {author} {\bibinfo {author} {\bibfnamefont {Junhu}\ \bibnamefont
  {Zhang}}, \bibinfo {author} {\bibfnamefont {Yunfeng}\ \bibnamefont {Li}},
  \bibinfo {author} {\bibfnamefont {Xuemin}\ \bibnamefont {Zhang}}, \ and\
  \bibinfo {author} {\bibfnamefont {Bai}\ \bibnamefont {Yang}},\ }\bibfield
  {title} {\enquote {\bibinfo {title} {Colloidal self-assembly meets
  nanofabrication: From two-dimensional colloidal crystals to nanostructure
  arrays},}\ }\href {\doibase 10.1002/adma.201000755} {\bibfield  {journal}
  {\bibinfo  {journal} {Adv. Mat.}\ }\textbf {\bibinfo {volume} {22}},\
  \bibinfo {pages} {4249--4269} (\bibinfo {year} {2010})}\BibitemShut {NoStop}%
\bibitem [{\citenamefont {Wu}\ \emph {et~al.}(2012)\citenamefont {Wu},
  \citenamefont {Xu}, \citenamefont {Sun}, \citenamefont {Fu}, \citenamefont
  {He},\ and\ \citenamefont {Matyjaszewski}}]{ReviewDesign2012}%
  \BibitemOpen
  \bibfield  {author} {\bibinfo {author} {\bibfnamefont {Dingcai}\ \bibnamefont
  {Wu}}, \bibinfo {author} {\bibfnamefont {Fei}\ \bibnamefont {Xu}}, \bibinfo
  {author} {\bibfnamefont {Bin}\ \bibnamefont {Sun}}, \bibinfo {author}
  {\bibfnamefont {Ruowen}\ \bibnamefont {Fu}}, \bibinfo {author} {\bibfnamefont
  {Hongkun}\ \bibnamefont {He}}, \ and\ \bibinfo {author} {\bibfnamefont
  {Krzysztof}\ \bibnamefont {Matyjaszewski}},\ }\bibfield  {title} {\enquote
  {\bibinfo {title} {Design and preparation of porous polymers},}\ }\href
  {\doibase 10.1021/cr200440z} {\bibfield  {journal} {\bibinfo  {journal}
  {Chem. Rev.}\ }\textbf {\bibinfo {volume} {112}},\ \bibinfo {pages}
  {3959--4015} (\bibinfo {year} {2012})}\BibitemShut {NoStop}%
\bibitem [{\citenamefont {Wang}\ \emph {et~al.}(2006)\citenamefont {Wang},
  \citenamefont {Adeyeye},\ and\ \citenamefont {Singh}}]{MagneticNano2006}%
  \BibitemOpen
  \bibfield  {author} {\bibinfo {author} {\bibfnamefont {C~C}\ \bibnamefont
  {Wang}}, \bibinfo {author} {\bibfnamefont {A~O}\ \bibnamefont {Adeyeye}}, \
  and\ \bibinfo {author} {\bibfnamefont {N}~\bibnamefont {Singh}},\ }\bibfield
  {title} {\enquote {\bibinfo {title} {Magnetic antidot nanostructures: effect
  of lattice geometry},}\ }\href {http://stacks.iop.org/0957-4484/17/i=6/a=015}
  {\bibfield  {journal} {\bibinfo  {journal} {Nanotechnology}\ }\textbf
  {\bibinfo {volume} {17}},\ \bibinfo {pages} {1629} (\bibinfo {year}
  {2006})}\BibitemShut {NoStop}%
\bibitem [{\citenamefont {Xia}\ \emph {et~al.}(2000)\citenamefont {Xia},
  \citenamefont {Gates}, \citenamefont {Yin},\ and\ \citenamefont
  {Lu}}]{ReviewColloids2000}%
  \BibitemOpen
  \bibfield  {author} {\bibinfo {author} {\bibfnamefont {Y.}~\bibnamefont
  {Xia}}, \bibinfo {author} {\bibfnamefont {B.}~\bibnamefont {Gates}}, \bibinfo
  {author} {\bibfnamefont {Y.}~\bibnamefont {Yin}}, \ and\ \bibinfo {author}
  {\bibfnamefont {Y.}~\bibnamefont {Lu}},\ }\bibfield  {title} {\enquote
  {\bibinfo {title} {Monodispersed colloidal spheres: Old materials with new
  applications},}\ }\href {\doibase
  10.1002/(SICI)1521-4095(200005)12:10<693::AID-ADMA693>3.0.CO;2-J} {\bibfield
  {journal} {\bibinfo  {journal} {Adv. Mat.}\ }\textbf {\bibinfo {volume}
  {12}},\ \bibinfo {pages} {693--713} (\bibinfo {year} {2000})}\BibitemShut
  {NoStop}%
\bibitem [{\citenamefont {Furumi}\ \emph {et~al.}(2010)\citenamefont {Furumi},
  \citenamefont {Fudouzi},\ and\ \citenamefont {Sawada}}]{ReviewCCPBG2010}%
  \BibitemOpen
  \bibfield  {author} {\bibinfo {author} {\bibfnamefont {Seuchi}\ \bibnamefont
  {Furumi}}, \bibinfo {author} {\bibfnamefont {Hiroshi}\ \bibnamefont
  {Fudouzi}}, \ and\ \bibinfo {author} {\bibfnamefont {Tsutomu}\ \bibnamefont
  {Sawada}},\ }\bibfield  {title} {\enquote {\bibinfo {title} {Self-organized
  colloidal crystals for photonics and laser applications},}\ }\href {\doibase
  10.1002/lpor.200910005} {\bibfield  {journal} {\bibinfo  {journal} {Laser \&
  Photon. Rev.}\ }\textbf {\bibinfo {volume} {4}},\ \bibinfo {pages} {205--220}
  (\bibinfo {year} {2010})}\BibitemShut {NoStop}%
\bibitem [{\citenamefont {von Freymann}\ \emph {et~al.}(2013)\citenamefont {von
  Freymann}, \citenamefont {Kitaev}, \citenamefont {Lotschz},\ and\
  \citenamefont {Ozin}}]{ReviewBottomSAPC2013}%
  \BibitemOpen
  \bibfield  {author} {\bibinfo {author} {\bibfnamefont {Georg}\ \bibnamefont
  {von Freymann}}, \bibinfo {author} {\bibfnamefont {Vladimir}\ \bibnamefont
  {Kitaev}}, \bibinfo {author} {\bibfnamefont {Bettina~V.}\ \bibnamefont
  {Lotschz}}, \ and\ \bibinfo {author} {\bibfnamefont {Geoffrey~A.}\
  \bibnamefont {Ozin}},\ }\bibfield  {title} {\enquote {\bibinfo {title}
  {Bottom-up assembly of photonic crystals},}\ }\href {\doibase
  10.1039/c2cs35309a} {\bibfield  {journal} {\bibinfo  {journal} {Chem. Soc.
  Rev.}\ }\textbf {\bibinfo {volume} {42}},\ \bibinfo {pages} {2528--2554}
  (\bibinfo {year} {2013})}\BibitemShut {NoStop}%
\bibitem [{\citenamefont {Murray}\ \emph {et~al.}(2000)\citenamefont {Murray},
  \citenamefont {Kagan},\ and\ \citenamefont {Bawendi}}]{murray2000synthesis}%
  \BibitemOpen
  \bibfield  {author} {\bibinfo {author} {\bibfnamefont {Christopher~B}\
  \bibnamefont {Murray}}, \bibinfo {author} {\bibfnamefont {CR}~\bibnamefont
  {Kagan}}, \ and\ \bibinfo {author} {\bibfnamefont {MG}~\bibnamefont
  {Bawendi}},\ }\bibfield  {title} {\enquote {\bibinfo {title} {Synthesis and
  characterization of monodisperse nanocrystals and close-packed nanocrystal
  assemblies},}\ }\href@noop {} {\bibfield  {journal} {\bibinfo  {journal}
  {Annu. Rev. Mater. Sci.}\ }\textbf {\bibinfo {volume} {30}},\ \bibinfo
  {pages} {545--610} (\bibinfo {year} {2000})}\BibitemShut {NoStop}%
\bibitem [{\citenamefont {Kim}\ \emph {et~al.}(2013)\citenamefont {Kim},
  \citenamefont {Voznyy}, \citenamefont {Zhitomirsky},\ and\ \citenamefont
  {Sargent}}]{QDReview2013}%
  \BibitemOpen
  \bibfield  {author} {\bibinfo {author} {\bibfnamefont {Jin~Young}\
  \bibnamefont {Kim}}, \bibinfo {author} {\bibfnamefont {Oleksandr}\
  \bibnamefont {Voznyy}}, \bibinfo {author} {\bibfnamefont {David}\
  \bibnamefont {Zhitomirsky}}, \ and\ \bibinfo {author} {\bibfnamefont
  {Edward~H.}\ \bibnamefont {Sargent}},\ }\bibfield  {title} {\enquote
  {\bibinfo {title} {25th anniversary article: Colloidal quantum dot materials
  and devices: A quarter-century of advances},}\ }\href@noop {} {\bibfield
  {journal} {\bibinfo  {journal} {Adv. Mat.}\ }\textbf {\bibinfo {volume}
  {25}},\ \bibinfo {pages} {4986--5010} (\bibinfo {year} {2013})}\BibitemShut
  {NoStop}%
\bibitem [{\citenamefont {Glotzer}\ and\ \citenamefont
  {Solomon}(2007)}]{glotzer2007anisotropy}%
  \BibitemOpen
  \bibfield  {author} {\bibinfo {author} {\bibfnamefont {S.C.}\ \bibnamefont
  {Glotzer}}\ and\ \bibinfo {author} {\bibfnamefont {M.J.}\ \bibnamefont
  {Solomon}},\ }\bibfield  {title} {\enquote {\bibinfo {title} {Anisotropy of
  building blocks and their assembly into complex structures},}\ }\href@noop {}
  {\bibfield  {journal} {\bibinfo  {journal} {Nat. Mater.}\ }\textbf {\bibinfo
  {volume} {6}},\ \bibinfo {pages} {557--562} (\bibinfo {year}
  {2007})}\BibitemShut {NoStop}%
\bibitem [{\citenamefont {Xia}\ \emph {et~al.}(2009)\citenamefont {Xia},
  \citenamefont {Xiong}, \citenamefont {Lim},\ and\ \citenamefont
  {Skrabalak}}]{ExpShapeNP2009}%
  \BibitemOpen
  \bibfield  {author} {\bibinfo {author} {\bibfnamefont {Younan}\ \bibnamefont
  {Xia}}, \bibinfo {author} {\bibfnamefont {Yujie}\ \bibnamefont {Xiong}},
  \bibinfo {author} {\bibfnamefont {Byungkwon}\ \bibnamefont {Lim}}, \ and\
  \bibinfo {author} {\bibfnamefont {Sara~E.}\ \bibnamefont {Skrabalak}},\
  }\bibfield  {title} {\enquote {\bibinfo {title} {Shape-controlled synthesis
  of metal nanocrystals: Simple chemistry meets complex physics?}}\ }\href
  {\doibase 10.1002/anie.200802248} {\bibfield  {journal} {\bibinfo  {journal}
  {Angew. Chem. Int. Edit.}\ }\textbf {\bibinfo {volume} {48}},\ \bibinfo
  {pages} {60--103} (\bibinfo {year} {2009})}\BibitemShut {NoStop}%
\bibitem [{\citenamefont {Damasceno}\ \emph {et~al.}(2012)\citenamefont
  {Damasceno}, \citenamefont {Engel},\ and\ \citenamefont
  {Glotzer}}]{Glotzerscience}%
  \BibitemOpen
  \bibfield  {author} {\bibinfo {author} {\bibfnamefont {Pablo~F.}\
  \bibnamefont {Damasceno}}, \bibinfo {author} {\bibfnamefont {Michael}\
  \bibnamefont {Engel}}, \ and\ \bibinfo {author} {\bibfnamefont {Sharon~C.}\
  \bibnamefont {Glotzer}},\ }\bibfield  {title} {\enquote {\bibinfo {title}
  {Predictive self-assembly of polyhedra into complex structures},}\ }\href
  {\doibase 10.1126/science.1220869} {\bibfield  {journal} {\bibinfo  {journal}
  {Science}\ }\textbf {\bibinfo {volume} {337}},\ \bibinfo {pages} {453--457}
  (\bibinfo {year} {2012})}\BibitemShut {NoStop}%
\bibitem [{\citenamefont {Agarwal}\ and\ \citenamefont
  {Escobedo}(2011)}]{EscobedoNatMat2011}%
  \BibitemOpen
  \bibfield  {author} {\bibinfo {author} {\bibfnamefont {Umang}\ \bibnamefont
  {Agarwal}}\ and\ \bibinfo {author} {\bibfnamefont {Fernando~A.}\ \bibnamefont
  {Escobedo}},\ }\bibfield  {title} {\enquote {\bibinfo {title} {Mesophase
  behaviour of polyhedral particles},}\ }\href {\doibase 10.1038/NMAT2959}
  {\bibfield  {journal} {\bibinfo  {journal} {Nat. Mater.}\ }\textbf {\bibinfo
  {volume} {10}},\ \bibinfo {pages} {230--235} (\bibinfo {year}
  {2011})}\BibitemShut {NoStop}%
\bibitem [{\citenamefont {Henzie}\ \emph {et~al.}(2012)\citenamefont {Henzie},
  \citenamefont {Gruenwald}, \citenamefont {Widmer-Cooper}, \citenamefont
  {Geissler},\ and\ \citenamefont {Yang}}]{GeisslerNatMat2012}%
  \BibitemOpen
  \bibfield  {author} {\bibinfo {author} {\bibfnamefont {Joel}\ \bibnamefont
  {Henzie}}, \bibinfo {author} {\bibfnamefont {Michael}\ \bibnamefont
  {Gruenwald}}, \bibinfo {author} {\bibfnamefont {Asaph}\ \bibnamefont
  {Widmer-Cooper}}, \bibinfo {author} {\bibfnamefont {Phillip~L.}\ \bibnamefont
  {Geissler}}, \ and\ \bibinfo {author} {\bibfnamefont {Peidong}\ \bibnamefont
  {Yang}},\ }\bibfield  {title} {\enquote {\bibinfo {title} {Self-assembly of
  uniform polyhedral silver nanocrystals into densest packings and exotic
  superlattices},}\ }\href {\doibase 10.1038/NMAT3178} {\bibfield  {journal}
  {\bibinfo  {journal} {Nat. Mater.}\ }\textbf {\bibinfo {volume} {11}},\
  \bibinfo {pages} {131--137} (\bibinfo {year} {2012})}\BibitemShut {NoStop}%
\bibitem [{\citenamefont {Quan}\ \emph {et~al.}(2014)\citenamefont {Quan},
  \citenamefont {Xu}, \citenamefont {Wang}, \citenamefont {Wen}, \citenamefont
  {Wang}, \citenamefont {Zhu}, \citenamefont {Li}, \citenamefont {Sheehan},
  \citenamefont {Wang}, \citenamefont {Smilgies}, \citenamefont {Luo},\ and\
  \citenamefont {Fang}}]{PtNanocubes}%
  \BibitemOpen
  \bibfield  {author} {\bibinfo {author} {\bibfnamefont {Zewei}\ \bibnamefont
  {Quan}}, \bibinfo {author} {\bibfnamefont {Hongwu}\ \bibnamefont {Xu}},
  \bibinfo {author} {\bibfnamefont {Chenyu}\ \bibnamefont {Wang}}, \bibinfo
  {author} {\bibfnamefont {Xiaodong}\ \bibnamefont {Wen}}, \bibinfo {author}
  {\bibfnamefont {Yuxuan}\ \bibnamefont {Wang}}, \bibinfo {author}
  {\bibfnamefont {Jinlong}\ \bibnamefont {Zhu}}, \bibinfo {author}
  {\bibfnamefont {Ruipeng}\ \bibnamefont {Li}}, \bibinfo {author}
  {\bibfnamefont {Chris~J.}\ \bibnamefont {Sheehan}}, \bibinfo {author}
  {\bibfnamefont {Zhongwu}\ \bibnamefont {Wang}}, \bibinfo {author}
  {\bibfnamefont {Detlef-M.}\ \bibnamefont {Smilgies}}, \bibinfo {author}
  {\bibfnamefont {Zhiping}\ \bibnamefont {Luo}}, \ and\ \bibinfo {author}
  {\bibfnamefont {Jiye}\ \bibnamefont {Fang}},\ }\bibfield  {title} {\enquote
  {\bibinfo {title} {Solvent-mediated self-assembly of nanocube
  superlattices},}\ }\href {\doibase 10.1021/ja408250q} {\bibfield  {journal}
  {\bibinfo  {journal} {J. Am. Chem. Soc.}\ }\textbf {\bibinfo {volume}
  {136}},\ \bibinfo {pages} {1352--1359} (\bibinfo {year} {2014})}\BibitemShut
  {NoStop}%
\bibitem [{\citenamefont {Sacanna}\ \emph {et~al.}(2013)\citenamefont
  {Sacanna}, \citenamefont {Pine},\ and\ \citenamefont {Yi}}]{RevShape2013}%
  \BibitemOpen
  \bibfield  {author} {\bibinfo {author} {\bibfnamefont {Stefano}\ \bibnamefont
  {Sacanna}}, \bibinfo {author} {\bibfnamefont {David~J.}\ \bibnamefont
  {Pine}}, \ and\ \bibinfo {author} {\bibfnamefont {Gi-Ra}\ \bibnamefont
  {Yi}},\ }\bibfield  {title} {\enquote {\bibinfo {title} {Engineering shape:
  the novel geometries of colloidal self-assembly},}\ }\href {\doibase
  10.1039/c3sm50500f} {\bibfield  {journal} {\bibinfo  {journal} {Soft Matter}\
  }\textbf {\bibinfo {volume} {9}},\ \bibinfo {pages} {8096--8106} (\bibinfo
  {year} {2013})}\BibitemShut {NoStop}%
\bibitem [{\citenamefont {Vega}\ and\ \citenamefont
  {Monson}(1998)}]{vega:9938}%
  \BibitemOpen
  \bibfield  {author} {\bibinfo {author} {\bibfnamefont {C.}~\bibnamefont
  {Vega}}\ and\ \bibinfo {author} {\bibfnamefont {P.~A.}\ \bibnamefont
  {Monson}},\ }\bibfield  {title} {\enquote {\bibinfo {title} {Solid--fluid
  equilibrium for a molecular model with short ranged directional forces},}\
  }\href {\doibase 10.1063/1.477660} {\bibfield  {journal} {\bibinfo  {journal}
  {J. Chem. Phys.}\ }\textbf {\bibinfo {volume} {109}},\ \bibinfo {pages}
  {9938--9949} (\bibinfo {year} {1998})}\BibitemShut {NoStop}%
\bibitem [{\citenamefont {Kraft}\ \emph {et~al.}(2012)\citenamefont {Kraft},
  \citenamefont {Ni}, \citenamefont {Smallenburg}, \citenamefont {Hermes},
  \citenamefont {Yoon}, \citenamefont {Weitz}, \citenamefont {van Blaaderen},
  \citenamefont {Groenewold}, \citenamefont {Dijkstra},\ and\ \citenamefont
  {Kegel}}]{SurfRoughPNAS2012}%
  \BibitemOpen
  \bibfield  {author} {\bibinfo {author} {\bibfnamefont {Daniela~J.}\
  \bibnamefont {Kraft}}, \bibinfo {author} {\bibfnamefont {Ran}\ \bibnamefont
  {Ni}}, \bibinfo {author} {\bibfnamefont {Frank}\ \bibnamefont {Smallenburg}},
  \bibinfo {author} {\bibfnamefont {Michiel}\ \bibnamefont {Hermes}}, \bibinfo
  {author} {\bibfnamefont {Kisun}\ \bibnamefont {Yoon}}, \bibinfo {author}
  {\bibfnamefont {David~A.}\ \bibnamefont {Weitz}}, \bibinfo {author}
  {\bibfnamefont {Alfons}\ \bibnamefont {van Blaaderen}}, \bibinfo {author}
  {\bibfnamefont {Jan}\ \bibnamefont {Groenewold}}, \bibinfo {author}
  {\bibfnamefont {Marjolein}\ \bibnamefont {Dijkstra}}, \ and\ \bibinfo
  {author} {\bibfnamefont {Willem~K.}\ \bibnamefont {Kegel}},\ }\bibfield
  {title} {\enquote {\bibinfo {title} {Surface roughness directed self-assembly
  of patchy particles into colloidal micelles},}\ }\href {\doibase
  10.1073/pnas.1116820109} {\bibfield  {journal} {\bibinfo  {journal} {Proc.
  Nat. Acad. Sci. USA}\ }\textbf {\bibinfo {volume} {109}},\ \bibinfo {pages}
  {10787--10792} (\bibinfo {year} {2012})}\BibitemShut {NoStop}%
\bibitem [{\citenamefont {Yi}\ \emph {et~al.}(2013)\citenamefont {Yi},
  \citenamefont {Pine},\ and\ \citenamefont {Sacanna}}]{ReviewpatchyMay2013}%
  \BibitemOpen
  \bibfield  {author} {\bibinfo {author} {\bibfnamefont {Gi-Ra}\ \bibnamefont
  {Yi}}, \bibinfo {author} {\bibfnamefont {David~J.}\ \bibnamefont {Pine}}, \
  and\ \bibinfo {author} {\bibfnamefont {Stefano}\ \bibnamefont {Sacanna}},\
  }\bibfield  {title} {\enquote {\bibinfo {title} {Recent progress on patchy
  colloids and their self-assembly},}\ }\href@noop {} {\bibfield  {journal}
  {\bibinfo  {journal} {J. Phys.:Condens. Matter}\ }\textbf {\bibinfo {volume}
  {25}} (\bibinfo {year} {2013})}\BibitemShut {NoStop}%
\bibitem [{\citenamefont {Jayaraman}(2013)}]{ArthiJayJPol2013}%
  \BibitemOpen
  \bibfield  {author} {\bibinfo {author} {\bibfnamefont {Arthi}\ \bibnamefont
  {Jayaraman}},\ }\bibfield  {title} {\enquote {\bibinfo {title} {Polymer
  grafted nanoparticles: Effect of chemical and physical heterogeneity in
  polymer grafts on particle assembly and dispersion},}\ }\href {\doibase
  10.1002/polb.23260} {\bibfield  {journal} {\bibinfo  {journal} {J. Polym.
  Sci. Part B: Polym. Phys.}\ }\textbf {\bibinfo {volume} {51}},\ \bibinfo
  {pages} {524--534} (\bibinfo {year} {2013})}\BibitemShut {NoStop}%
\bibitem [{\citenamefont {Courty}\ \emph {et~al.}(2011)\citenamefont {Courty},
  \citenamefont {Richardi}, \citenamefont {Albouy},\ and\ \citenamefont
  {Pileni}}]{MPPileni2011}%
  \BibitemOpen
  \bibfield  {author} {\bibinfo {author} {\bibfnamefont {Alexa}\ \bibnamefont
  {Courty}}, \bibinfo {author} {\bibfnamefont {Johannes}\ \bibnamefont
  {Richardi}}, \bibinfo {author} {\bibfnamefont {Pierre-Antoine}\ \bibnamefont
  {Albouy}}, \ and\ \bibinfo {author} {\bibfnamefont {Marie-Paule}\
  \bibnamefont {Pileni}},\ }\bibfield  {title} {\enquote {\bibinfo {title} {How
  to control the crystalline structure of supracrystals of 5-nm silver
  nanocrystals},}\ }\href@noop {} {\bibfield  {journal} {\bibinfo  {journal}
  {Chem. Mater.}\ }\textbf {\bibinfo {volume} {23}},\ \bibinfo {pages}
  {4186--4192} (\bibinfo {year} {2011})}\BibitemShut {NoStop}%
\bibitem [{\citenamefont {Shevchenko}\ \emph {et~al.}(2006)\citenamefont
  {Shevchenko}, \citenamefont {Talapin}, \citenamefont {Kotov}, \citenamefont
  {O'Brien},\ and\ \citenamefont {Murray}}]{shevchenko2006structural}%
  \BibitemOpen
  \bibfield  {author} {\bibinfo {author} {\bibfnamefont {Elena~V}\ \bibnamefont
  {Shevchenko}}, \bibinfo {author} {\bibfnamefont {Dmitri~V}\ \bibnamefont
  {Talapin}}, \bibinfo {author} {\bibfnamefont {Nicholas~A}\ \bibnamefont
  {Kotov}}, \bibinfo {author} {\bibfnamefont {Stephen}\ \bibnamefont
  {O'Brien}}, \ and\ \bibinfo {author} {\bibfnamefont {Christopher~B}\
  \bibnamefont {Murray}},\ }\bibfield  {title} {\enquote {\bibinfo {title}
  {Structural diversity in binary nanoparticle superlattices},}\ }\href@noop {}
  {\bibfield  {journal} {\bibinfo  {journal} {Nature (London)}\ }\textbf
  {\bibinfo {volume} {439}},\ \bibinfo {pages} {55--59} (\bibinfo {year}
  {2006})}\BibitemShut {NoStop}%
\bibitem [{\citenamefont {Pavan}\ \emph {et~al.}(2012)\citenamefont {Pavan},
  \citenamefont {Ploshnik},\ and\ \citenamefont {Shenhar}}]{royshenhar2012}%
  \BibitemOpen
  \bibfield  {author} {\bibinfo {author} {\bibfnamefont {Mariela~J.}\
  \bibnamefont {Pavan}}, \bibinfo {author} {\bibfnamefont {Elina}\ \bibnamefont
  {Ploshnik}}, \ and\ \bibinfo {author} {\bibfnamefont {Roy}\ \bibnamefont
  {Shenhar}},\ }\bibfield  {title} {\enquote {\bibinfo {title} {Nanoparticle
  assembly on topographical polymer templates: Effects of spin rate,
  nanoparticle size, ligand, and concentration},}\ }\href {\doibase
  10.1021/jp308910w} {\bibfield  {journal} {\bibinfo  {journal} {J. Phys. Chem.
  B}\ }\textbf {\bibinfo {volume} {116}},\ \bibinfo {pages} {13922--13931}
  (\bibinfo {year} {2012})}\BibitemShut {NoStop}%
\bibitem [{\citenamefont {Kinge}\ \emph {et~al.}(2008)\citenamefont {Kinge},
  \citenamefont {Crego-Calama},\ and\ \citenamefont
  {Reinhoudt}}]{CPC-NP-Surfaces-Interfaces}%
  \BibitemOpen
  \bibfield  {author} {\bibinfo {author} {\bibfnamefont {Sachin}\ \bibnamefont
  {Kinge}}, \bibinfo {author} {\bibfnamefont {Mercedes}\ \bibnamefont
  {Crego-Calama}}, \ and\ \bibinfo {author} {\bibfnamefont {David~N.}\
  \bibnamefont {Reinhoudt}},\ }\bibfield  {title} {\enquote {\bibinfo {title}
  {Self-assembling nanoparticles at surfaces and interfaces},}\ }\href
  {\doibase 10.1002/cphc.200700475} {\bibfield  {journal} {\bibinfo  {journal}
  {Chem. Phys. Chem.}\ }\textbf {\bibinfo {volume} {9}},\ \bibinfo {pages}
  {20--42} (\bibinfo {year} {2008})}\BibitemShut {NoStop}%
\bibitem [{\citenamefont {Ho}\ \emph {et~al.}(1990)\citenamefont {Ho},
  \citenamefont {Chan},\ and\ \citenamefont {Soukoulis}}]{DIAphoton}%
  \BibitemOpen
  \bibfield  {author} {\bibinfo {author} {\bibfnamefont {K.~M.}\ \bibnamefont
  {Ho}}, \bibinfo {author} {\bibfnamefont {C.~T.}\ \bibnamefont {Chan}}, \ and\
  \bibinfo {author} {\bibfnamefont {C.~M.}\ \bibnamefont {Soukoulis}},\
  }\bibfield  {title} {\enquote {\bibinfo {title} {Existence of a photonic gap
  in periodic dielectric structures},}\ }\href {\doibase
  10.1103/PhysRevLett.65.3152} {\bibfield  {journal} {\bibinfo  {journal}
  {Phys. Rev. Lett.}\ }\textbf {\bibinfo {volume} {65}},\ \bibinfo {pages}
  {3152--3155} (\bibinfo {year} {1990})}\BibitemShut {NoStop}%
\bibitem [{\citenamefont {S\"{o}z\"{u}er}\ and\ \citenamefont
  {Haus}(1993)}]{PCsimplecubic}%
  \BibitemOpen
  \bibfield  {author} {\bibinfo {author} {\bibfnamefont {H.~Sami}\ \bibnamefont
  {S\"{o}z\"{u}er}}\ and\ \bibinfo {author} {\bibfnamefont {Joseph~W.}\
  \bibnamefont {Haus}},\ }\bibfield  {title} {\enquote {\bibinfo {title}
  {Photonic bands: Simple-cubic lattice},}\ }\href {\doibase
  10.1364/JOSAB.10.000296} {\bibfield  {journal} {\bibinfo  {journal} {J. Opt.
  Soc. Am. B}\ }\textbf {\bibinfo {volume} {10}},\ \bibinfo {pages} {296--302}
  (\bibinfo {year} {1993})}\BibitemShut {NoStop}%
\bibitem [{\citenamefont {Yoshida}\ and\ \citenamefont
  {Kamakura}(1972)}]{YKorig1}%
  \BibitemOpen
  \bibfield  {author} {\bibinfo {author} {\bibfnamefont {Takeshi}\ \bibnamefont
  {Yoshida}}\ and\ \bibinfo {author} {\bibfnamefont {Shiro}\ \bibnamefont
  {Kamakura}},\ }\bibfield  {title} {\enquote {\bibinfo {title} {Theory of
  melting at high pressures. {II}},}\ }\href {\doibase 10.1143/PTP.47.1801}
  {\bibfield  {journal} {\bibinfo  {journal} {Prog. Theor. Phys.}\ }\textbf
  {\bibinfo {volume} {47}},\ \bibinfo {pages} {1801--1816} (\bibinfo {year}
  {1972})}\BibitemShut {NoStop}%
\bibitem [{\citenamefont {Fomin}\ \emph {et~al.}(2008)\citenamefont {Fomin},
  \citenamefont {Gribova}, \citenamefont {Ryzhov}, \citenamefont {Stishov},\
  and\ \citenamefont {Frenkel}}]{fominsc}%
  \BibitemOpen
  \bibfield  {author} {\bibinfo {author} {\bibfnamefont {Yu.~D.}\ \bibnamefont
  {Fomin}}, \bibinfo {author} {\bibfnamefont {N.~V.}\ \bibnamefont {Gribova}},
  \bibinfo {author} {\bibfnamefont {V.~N.}\ \bibnamefont {Ryzhov}}, \bibinfo
  {author} {\bibfnamefont {S.~M.}\ \bibnamefont {Stishov}}, \ and\ \bibinfo
  {author} {\bibfnamefont {Daan}\ \bibnamefont {Frenkel}},\ }\bibfield  {title}
  {\enquote {\bibinfo {title} {Quasibinary amorphous phase in a
  three-dimensional system of particles with repulsive-shoulder
  interactions},}\ }\href {\doibase 10.1063/1.2965880} {\bibfield  {journal}
  {\bibinfo  {journal} {J. Chem. Phys.}\ }\textbf {\bibinfo {volume} {129}},\
  \bibinfo {eid} {064512} (\bibinfo {year} {2008})}\BibitemShut {NoStop}%
\bibitem [{\citenamefont {Marcotte}\ \emph {et~al.}(2013)\citenamefont
  {Marcotte}, \citenamefont {Stillinger},\ and\ \citenamefont
  {Torquato}}]{marcottediamondpaper}%
  \BibitemOpen
  \bibfield  {author} {\bibinfo {author} {\bibfnamefont {\'{E}.}\ \bibnamefont
  {Marcotte}}, \bibinfo {author} {\bibfnamefont {F.~H.}\ \bibnamefont
  {Stillinger}}, \ and\ \bibinfo {author} {\bibfnamefont {Salvatore}\
  \bibnamefont {Torquato}},\ }\bibfield  {title} {\enquote {\bibinfo {title}
  {Communication: Designed diamond ground state via optimized isotropic
  monotonic pair potentials},}\ }\href {\doibase 10.1063/1.4790634} {\bibfield
  {journal} {\bibinfo  {journal} {J. Chem. Phys.}\ }\textbf {\bibinfo {volume}
  {138}},\ \bibinfo {eid} {061101} (\bibinfo {year} {2013})}\BibitemShut
  {NoStop}%
\bibitem [{\citenamefont {Edlund}\ \emph {et~al.}(2013)\citenamefont {Edlund},
  \citenamefont {Lindgren},\ and\ \citenamefont {Jacobi}}]{edlundjcp2013}%
  \BibitemOpen
  \bibfield  {author} {\bibinfo {author} {\bibfnamefont {E.}~\bibnamefont
  {Edlund}}, \bibinfo {author} {\bibfnamefont {O.}~\bibnamefont {Lindgren}}, \
  and\ \bibinfo {author} {\bibfnamefont {M.~Nilsson}\ \bibnamefont {Jacobi}},\
  }\bibfield  {title} {\enquote {\bibinfo {title} {Using the uncertainty
  principle to design simple interactions for targeted self-assembly},}\ }\href
  {\doibase 10.1063/1.4812727} {\bibfield  {journal} {\bibinfo  {journal} {J.
  Chem. Phys.}\ }\textbf {\bibinfo {volume} {139}},\ \bibinfo {eid} {024107}
  (\bibinfo {year} {2013})}\BibitemShut {NoStop}%
\bibitem [{\citenamefont {Jain}\ \emph
  {et~al.}(2013{\natexlab{a}})\citenamefont {Jain}, \citenamefont {Errington},\
  and\ \citenamefont {Truskett}}]{SMJainTMT2013}%
  \BibitemOpen
  \bibfield  {author} {\bibinfo {author} {\bibfnamefont {Avni}\ \bibnamefont
  {Jain}}, \bibinfo {author} {\bibfnamefont {Jeffrey~R.}\ \bibnamefont
  {Errington}}, \ and\ \bibinfo {author} {\bibfnamefont {Thomas~M.}\
  \bibnamefont {Truskett}},\ }\bibfield  {title} {\enquote {\bibinfo {title}
  {Inverse design of simple pairwise interactions with low-coordinated {3D}
  lattice ground states},}\ }\href {\doibase 10.1039/C3SM27785B} {\bibfield
  {journal} {\bibinfo  {journal} {Soft Matter}\ }\textbf {\bibinfo {volume}
  {9}},\ \bibinfo {pages} {3866--3870} (\bibinfo {year}
  {2013}{\natexlab{a}})}\BibitemShut {NoStop}%
\bibitem [{\citenamefont {P\`{a}mies}\ \emph {et~al.}(2009)\citenamefont
  {P\`{a}mies}, \citenamefont {Cacciuto},\ and\ \citenamefont
  {Frenkel}}]{hertz}%
  \BibitemOpen
  \bibfield  {author} {\bibinfo {author} {\bibfnamefont {Josep~C.}\
  \bibnamefont {P\`{a}mies}}, \bibinfo {author} {\bibfnamefont {Angelo}\
  \bibnamefont {Cacciuto}}, \ and\ \bibinfo {author} {\bibfnamefont {Daan}\
  \bibnamefont {Frenkel}},\ }\bibfield  {title} {\enquote {\bibinfo {title}
  {Phase diagram of hertzian spheres},}\ }\href {\doibase 10.1063/1.3186742}
  {\bibfield  {journal} {\bibinfo  {journal} {J. Chem. Phys.}\ }\textbf
  {\bibinfo {volume} {131}},\ \bibinfo {eid} {044514} (\bibinfo {year}
  {2009})}\BibitemShut {NoStop}%
\bibitem [{\citenamefont {Watzlawek}\ \emph {et~al.}(1999)\citenamefont
  {Watzlawek}, \citenamefont {Likos},\ and\ \citenamefont
  {L\"owen}}]{LikosStarPot}%
  \BibitemOpen
  \bibfield  {author} {\bibinfo {author} {\bibfnamefont {M.}~\bibnamefont
  {Watzlawek}}, \bibinfo {author} {\bibfnamefont {C.~N.}\ \bibnamefont
  {Likos}}, \ and\ \bibinfo {author} {\bibfnamefont {H.}~\bibnamefont
  {L\"owen}},\ }\bibfield  {title} {\enquote {\bibinfo {title} {Phase diagram
  of star polymer solutions},}\ }\href@noop {} {\bibfield  {journal} {\bibinfo
  {journal} {Phys. Rev. Lett.}\ }\textbf {\bibinfo {volume} {82}},\ \bibinfo
  {pages} {5289--5292} (\bibinfo {year} {1999})}\BibitemShut {NoStop}%
\bibitem [{\citenamefont {Jagla}(1999)}]{EAJagla1999}%
  \BibitemOpen
  \bibfield  {author} {\bibinfo {author} {\bibfnamefont {E.~A.}\ \bibnamefont
  {Jagla}},\ }\bibfield  {title} {\enquote {\bibinfo {title} {Minimum energy
  configurations of repelling particles in two dimensions},}\ }\href@noop {}
  {\bibfield  {journal} {\bibinfo  {journal} {J. Chem. Phys.}\ }\textbf
  {\bibinfo {volume} {110}} (\bibinfo {year} {1999})}\BibitemShut {NoStop}%
\bibitem [{\citenamefont {Camp}(2003)}]{PJCamp2003}%
  \BibitemOpen
  \bibfield  {author} {\bibinfo {author} {\bibfnamefont {Philip~J.}\
  \bibnamefont {Camp}},\ }\bibfield  {title} {\enquote {\bibinfo {title}
  {Structure and phase behavior of a two-dimensional system with core-softened
  and long-range repulsive interactions},}\ }\href {\doibase
  10.1103/PhysRevE.68.061506} {\bibfield  {journal} {\bibinfo  {journal} {Phys.
  Rev. E}\ }\textbf {\bibinfo {volume} {68}},\ \bibinfo {pages} {061506}
  (\bibinfo {year} {2003})}\BibitemShut {NoStop}%
\bibitem [{\citenamefont {Malescio}\ and\ \citenamefont
  {Pellicane}(2003)}]{GMalescio2003}%
  \BibitemOpen
  \bibfield  {author} {\bibinfo {author} {\bibfnamefont {G}~\bibnamefont
  {Malescio}}\ and\ \bibinfo {author} {\bibfnamefont {G}~\bibnamefont
  {Pellicane}},\ }\bibfield  {title} {\enquote {\bibinfo {title} {Stripe phases
  from isotropic repulsive interactions},}\ }\href {\doibase 10.1038/nmat820}
  {\bibfield  {journal} {\bibinfo  {journal} {Nat. Mater.}\ }\textbf {\bibinfo
  {volume} {2}},\ \bibinfo {pages} {97--100} (\bibinfo {year}
  {2003})}\BibitemShut {NoStop}%
\bibitem [{\citenamefont {Marcotte}\ \emph
  {et~al.}(2011{\natexlab{a}})\citenamefont {Marcotte}, \citenamefont
  {Stillinger},\ and\ \citenamefont {Torquato}}]{JCP2dmonotonic}%
  \BibitemOpen
  \bibfield  {author} {\bibinfo {author} {\bibfnamefont {\'{E}.}\ \bibnamefont
  {Marcotte}}, \bibinfo {author} {\bibfnamefont {F.~H.}\ \bibnamefont
  {Stillinger}}, \ and\ \bibinfo {author} {\bibfnamefont {S.}~\bibnamefont
  {Torquato}},\ }\bibfield  {title} {\enquote {\bibinfo {title} {Unusual ground
  states via monotonic convex pair potentials},}\ }\href {\doibase
  10.1063/1.3576141} {\bibfield  {journal} {\bibinfo  {journal} {J. Chem.
  Phys.}\ }\textbf {\bibinfo {volume} {134}},\ \bibinfo {eid} {164105}
  (\bibinfo {year} {2011}{\natexlab{a}})}\BibitemShut {NoStop}%
\bibitem [{\citenamefont {Zhang}\ \emph {et~al.}(2013)\citenamefont {Zhang},
  \citenamefont {Stillinger},\ and\ \citenamefont
  {Torquato}}]{ZhangTorquato2013}%
  \BibitemOpen
  \bibfield  {author} {\bibinfo {author} {\bibfnamefont {G.}~\bibnamefont
  {Zhang}}, \bibinfo {author} {\bibfnamefont {F.~H.}\ \bibnamefont
  {Stillinger}}, \ and\ \bibinfo {author} {\bibfnamefont {S.}~\bibnamefont
  {Torquato}},\ }\bibfield  {title} {\enquote {\bibinfo {title} {Probing the
  limitations of isotropic pair potentials to produce ground-state structural
  extremes via inverse statistical mechanics},}\ }\href@noop {} {\bibfield
  {journal} {\bibinfo  {journal} {Phys. Rev. E}\ }\textbf {\bibinfo {volume}
  {88}},\ \bibinfo {pages} {042309} (\bibinfo {year} {2013})}\BibitemShut
  {NoStop}%
\bibitem [{\citenamefont {Lafitte}\ \emph {et~al.}(2014)\citenamefont
  {Lafitte}, \citenamefont {Kumar},\ and\ \citenamefont
  {Panagiotopoulos}}]{AGP-SK-SM-2014}%
  \BibitemOpen
  \bibfield  {author} {\bibinfo {author} {\bibfnamefont {Thomas}\ \bibnamefont
  {Lafitte}}, \bibinfo {author} {\bibfnamefont {Sanat~K.}\ \bibnamefont
  {Kumar}}, \ and\ \bibinfo {author} {\bibfnamefont {Athanassios~Z.}\
  \bibnamefont {Panagiotopoulos}},\ }\bibfield  {title} {\enquote {\bibinfo
  {title} {Self-assembly of polymer-grafted nanoparticles in thin films},}\
  }\href {\doibase 10.1039/C3SM52328D} {\bibfield  {journal} {\bibinfo
  {journal} {Soft Matter}\ }\textbf {\bibinfo {volume} {10}},\ \bibinfo {pages}
  {786--794} (\bibinfo {year} {2014})}\BibitemShut {NoStop}%
\bibitem [{\citenamefont {Dudalov}\ \emph {et~al.}(2014)\citenamefont
  {Dudalov}, \citenamefont {Fomin}, \citenamefont {Tsiok},\ and\ \citenamefont
  {Ryzhov}}]{PRL2Dfomin2014}%
  \BibitemOpen
  \bibfield  {author} {\bibinfo {author} {\bibfnamefont {D.E.}\ \bibnamefont
  {Dudalov}}, \bibinfo {author} {\bibfnamefont {Yu.D.}\ \bibnamefont {Fomin}},
  \bibinfo {author} {\bibfnamefont {E.N.}\ \bibnamefont {Tsiok}}, \ and\
  \bibinfo {author} {\bibfnamefont {V.N.}\ \bibnamefont {Ryzhov}},\ }\bibfield
  {title} {\enquote {\bibinfo {title} {Anomalous melting scenario of the
  two-dimensional core-softened system},}\ }\href {\doibase
  10.1103/PhysRevLett.112.157803} {\bibfield  {journal} {\bibinfo  {journal}
  {Phys. Rev. Lett.}\ }\textbf {\bibinfo {volume} {112}},\ \bibinfo {pages}
  {157803} (\bibinfo {year} {2014})}\BibitemShut {NoStop}%
\bibitem [{\citenamefont {Osterman}\ \emph {et~al.}(2007)\citenamefont
  {Osterman}, \citenamefont {Babi\ifmmode~\check{c}\else \v{c}\fi{}},
  \citenamefont {Poberaj}, \citenamefont {Dobnikar},\ and\ \citenamefont
  {Ziherl}}]{PZiherlPRL2007}%
  \BibitemOpen
  \bibfield  {author} {\bibinfo {author} {\bibfnamefont {N.}~\bibnamefont
  {Osterman}}, \bibinfo {author} {\bibfnamefont {D.}~\bibnamefont
  {Babi\ifmmode~\check{c}\else \v{c}\fi{}}}, \bibinfo {author} {\bibfnamefont
  {I.}~\bibnamefont {Poberaj}}, \bibinfo {author} {\bibfnamefont
  {J.}~\bibnamefont {Dobnikar}}, \ and\ \bibinfo {author} {\bibfnamefont
  {P.}~\bibnamefont {Ziherl}},\ }\bibfield  {title} {\enquote {\bibinfo {title}
  {Observation of condensed phases of quasiplanar core-softened colloids},}\
  }\href {\doibase 10.1103/PhysRevLett.99.248301} {\bibfield  {journal}
  {\bibinfo  {journal} {Phys. Rev. Lett.}\ }\textbf {\bibinfo {volume} {99}},\
  \bibinfo {pages} {248301} (\bibinfo {year} {2007})}\BibitemShut {NoStop}%
\bibitem [{\citenamefont {Chen}\ \emph {et~al.}(2011)\citenamefont {Chen},
  \citenamefont {Bae},\ and\ \citenamefont {Granick}}]{SteveGNature}%
  \BibitemOpen
  \bibfield  {author} {\bibinfo {author} {\bibfnamefont {Qian}\ \bibnamefont
  {Chen}}, \bibinfo {author} {\bibfnamefont {Sung~Chul}\ \bibnamefont {Bae}}, \
  and\ \bibinfo {author} {\bibfnamefont {Steve}\ \bibnamefont {Granick}},\
  }\bibfield  {title} {\enquote {\bibinfo {title} {Directed self-assembly of a
  colloidal kagome lattice},}\ }\href@noop {} {\bibfield  {journal} {\bibinfo
  {journal} {Nature (London)}\ }\textbf {\bibinfo {volume} {469}},\ \bibinfo
  {pages} {381--384} (\bibinfo {year} {2011})}\BibitemShut {NoStop}%
\bibitem [{\citenamefont {Evers}\ \emph {et~al.}(2013)\citenamefont {Evers},
  \citenamefont {Goris}, \citenamefont {Bals}, \citenamefont {Casavola},
  \citenamefont {de~Graaf}, \citenamefont {Roij}, \citenamefont {Dijkstra},\
  and\ \citenamefont {Vanmaekelbergh}}]{WHEvers2012}%
  \BibitemOpen
  \bibfield  {author} {\bibinfo {author} {\bibfnamefont {Wiel~H.}\ \bibnamefont
  {Evers}}, \bibinfo {author} {\bibfnamefont {Bart}\ \bibnamefont {Goris}},
  \bibinfo {author} {\bibfnamefont {Sara}\ \bibnamefont {Bals}}, \bibinfo
  {author} {\bibfnamefont {Marianna}\ \bibnamefont {Casavola}}, \bibinfo
  {author} {\bibfnamefont {Joost}\ \bibnamefont {de~Graaf}}, \bibinfo {author}
  {\bibfnamefont {Ren{\'{e}}~van}\ \bibnamefont {Roij}}, \bibinfo {author}
  {\bibfnamefont {Marjolein}\ \bibnamefont {Dijkstra}}, \ and\ \bibinfo
  {author} {\bibfnamefont {Dani{\"{e}}l}\ \bibnamefont {Vanmaekelbergh}},\
  }\bibfield  {title} {\enquote {\bibinfo {title} {Low-dimensional
  semiconductor superlattices formed by geometric control over nanocrystal
  attachment},}\ }\href {\doibase 10.1021/nl303322k} {\bibfield  {journal}
  {\bibinfo  {journal} {Nano Lett.}\ }\textbf {\bibinfo {volume} {13}},\
  \bibinfo {pages} {2317--2323} (\bibinfo {year} {2013})}\BibitemShut {NoStop}%
\bibitem [{\citenamefont {Antlanger}\ \emph {et~al.}(2011)\citenamefont
  {Antlanger}, \citenamefont {Doppelbauer},\ and\ \citenamefont
  {Kahl}}]{2DPatchydesign}%
  \BibitemOpen
  \bibfield  {author} {\bibinfo {author} {\bibfnamefont {Moritz}\ \bibnamefont
  {Antlanger}}, \bibinfo {author} {\bibfnamefont {G{\"{u}}nther}\ \bibnamefont
  {Doppelbauer}}, \ and\ \bibinfo {author} {\bibfnamefont {Gerhard}\
  \bibnamefont {Kahl}},\ }\bibfield  {title} {\enquote {\bibinfo {title} {On
  the stability of {A}rchimedean tilings formed by patchy particles},}\ }\href
  {http://stacks.iop.org/0953-8984/23/i=40/a=404206} {\bibfield  {journal}
  {\bibinfo  {journal} {J. Phys.: Condens. Matter}\ }\textbf {\bibinfo {volume}
  {23}},\ \bibinfo {pages} {404206} (\bibinfo {year} {2011})}\BibitemShut
  {NoStop}%
\bibitem [{\citenamefont {Millan}\ \emph {et~al.}(2014)\citenamefont {Millan},
  \citenamefont {Ortiz}, \citenamefont {van Anders},\ and\ \citenamefont
  {Glotzer}}]{ATselfassembly2014}%
  \BibitemOpen
  \bibfield  {author} {\bibinfo {author} {\bibfnamefont {Jaime~A.}\
  \bibnamefont {Millan}}, \bibinfo {author} {\bibfnamefont {Daniel}\
  \bibnamefont {Ortiz}}, \bibinfo {author} {\bibfnamefont {Greg}\ \bibnamefont
  {van Anders}}, \ and\ \bibinfo {author} {\bibfnamefont {Sharon~C.}\
  \bibnamefont {Glotzer}},\ }\bibfield  {title} {\enquote {\bibinfo {title}
  {Self-assembly of {A}rchimedean tilings with enthalpically and entropically
  patchy polygons},}\ }\href {\doibase 10.1021/nn500147u} {\bibfield  {journal}
  {\bibinfo  {journal} {ACS Nano}\ }\textbf {\bibinfo {volume} {8}},\ \bibinfo
  {pages} {2918--2928} (\bibinfo {year} {2014})}\BibitemShut {NoStop}%
\bibitem [{\citenamefont {Bianchi}\ \emph {et~al.}(2011)\citenamefont
  {Bianchi}, \citenamefont {Blaak},\ and\ \citenamefont
  {Likos}}]{PCCPPatchy-perspective}%
  \BibitemOpen
  \bibfield  {author} {\bibinfo {author} {\bibfnamefont {Emanuela}\
  \bibnamefont {Bianchi}}, \bibinfo {author} {\bibfnamefont {Ronald}\
  \bibnamefont {Blaak}}, \ and\ \bibinfo {author} {\bibfnamefont {Christos~N.}\
  \bibnamefont {Likos}},\ }\bibfield  {title} {\enquote {\bibinfo {title}
  {Patchy colloids: state of the art and perspectives},}\ }\href {\doibase
  10.1039/C0CP02296A} {\bibfield  {journal} {\bibinfo  {journal} {Phys. Chem.
  Chem. Phys.}\ }\textbf {\bibinfo {volume} {13}},\ \bibinfo {pages}
  {6397--6410} (\bibinfo {year} {2011})}\BibitemShut {NoStop}%
\bibitem [{\citenamefont {Romano}\ and\ \citenamefont
  {Sciortino}(2012)}]{sciortinonatcom}%
  \BibitemOpen
  \bibfield  {author} {\bibinfo {author} {\bibfnamefont {Flavio}\ \bibnamefont
  {Romano}}\ and\ \bibinfo {author} {\bibfnamefont {Francesco}\ \bibnamefont
  {Sciortino}},\ }\bibfield  {title} {\enquote {\bibinfo {title} {Patterning
  symmetry in the rational design of colloidal crystals},}\ }\href@noop {}
  {\bibfield  {journal} {\bibinfo  {journal} {Nat. Commun.}\ }\textbf {\bibinfo
  {volume} {3}},\ \bibinfo {pages} {975} (\bibinfo {year} {2012})}\BibitemShut
  {NoStop}%
\bibitem [{Note1()}]{Note1}%
  \BibitemOpen
  \bibinfo {note} {The term `analogous structures' refers to the pairs of 2D-3D
  lattices (e.g., honeycomb-diamond and square-simple cubic) that have specific
  coordination-shell similarities that can allow a single isotropic pair
  potential to favor the stability of both. This structural similarity is
  addressed both in the discussion of Fig.~3 and Fig.~S1 (in the Supplementary
  Information)}\BibitemShut {NoStop}%
\bibitem [{\citenamefont {Torquato}(2009)}]{TorquatoRev}%
  \BibitemOpen
  \bibfield  {author} {\bibinfo {author} {\bibfnamefont {Salvatore}\
  \bibnamefont {Torquato}},\ }\bibfield  {title} {\enquote {\bibinfo {title}
  {Inverse optimization techniques for targeted self-assembly},}\ }\href
  {\doibase 10.1039/B814211B} {\bibfield  {journal} {\bibinfo  {journal} {Soft
  Matter}\ }\textbf {\bibinfo {volume} {5}},\ \bibinfo {pages} {1157--1173}
  (\bibinfo {year} {2009})}\BibitemShut {NoStop}%
\bibitem [{\citenamefont {Jain}\ \emph {et~al.}()\citenamefont {Jain},
  \citenamefont {Bollinger},\ and\ \citenamefont {Truskett}}]{AIChEPersp}%
  \BibitemOpen
  \bibfield  {author} {\bibinfo {author} {\bibfnamefont {Avni}\ \bibnamefont
  {Jain}}, \bibinfo {author} {\bibfnamefont {Jonathan~A.}\ \bibnamefont
  {Bollinger}}, \ and\ \bibinfo {author} {\bibfnamefont {Thomas~M.}\
  \bibnamefont {Truskett}},\ }\bibfield  {title} {\enquote {\bibinfo {title}
  {Perspective: Inverse methods for material design},}\ }\href {\doibase
  10.1002/aic.14491} {\bibfield  {journal} {\bibinfo  {journal} {AIChE J.}\
  }10.1002/aic.14491}\BibitemShut {NoStop}%
\bibitem [{Note2()}]{Note2}%
  \BibitemOpen
  \bibinfo {note} {We note that particles confined to a 2D monolayer, such as
  at a liquid-liquid interface or on a substrate, may interact via an effective
  pair potential that is different from the one that the same particles
  experience in a 3D bulk fluid.}\BibitemShut {Stop}%
\bibitem [{\citenamefont {Schapotschnikow}\ \emph {et~al.}(2008)\citenamefont
  {Schapotschnikow}, \citenamefont {Pool},\ and\ \citenamefont
  {Vlugt}}]{schapotschnikow2008molecular}%
  \BibitemOpen
  \bibfield  {author} {\bibinfo {author} {\bibfnamefont {Philipp}\ \bibnamefont
  {Schapotschnikow}}, \bibinfo {author} {\bibfnamefont {René}\ \bibnamefont
  {Pool}}, \ and\ \bibinfo {author} {\bibfnamefont {Thijs~JH}\ \bibnamefont
  {Vlugt}},\ }\bibfield  {title} {\enquote {\bibinfo {title} {Molecular
  simulations of interacting nanocrystals},}\ }\href@noop {} {\bibfield
  {journal} {\bibinfo  {journal} {Nano Lett.}\ }\textbf {\bibinfo {volume}
  {8}},\ \bibinfo {pages} {2930--2934} (\bibinfo {year} {2008})}\BibitemShut
  {NoStop}%
\bibitem [{\citenamefont {Gottwald}\ \emph {et~al.}(2005)\citenamefont
  {Gottwald}, \citenamefont {Kahl},\ and\ \citenamefont {Likos}}]{GALikos1}%
  \BibitemOpen
  \bibfield  {author} {\bibinfo {author} {\bibfnamefont {Dieter}\ \bibnamefont
  {Gottwald}}, \bibinfo {author} {\bibfnamefont {Gerhard}\ \bibnamefont
  {Kahl}}, \ and\ \bibinfo {author} {\bibfnamefont {Christos~N.}\ \bibnamefont
  {Likos}},\ }\bibfield  {title} {\enquote {\bibinfo {title} {Predicting
  equilibrium structures in freezing processes.}}\ }\href@noop {} {\bibfield
  {journal} {\bibinfo  {journal} {J. Chem. Phys.}\ }\textbf {\bibinfo {volume}
  {122}},\ \bibinfo {pages} {204503} (\bibinfo {year} {2005})}\BibitemShut
  {NoStop}%
\bibitem [{\citenamefont {Bianchi}\ \emph {et~al.}(2012)\citenamefont
  {Bianchi}, \citenamefont {Doppelbauer}, \citenamefont {Filion}, \citenamefont
  {Dijkstra},\ and\ \citenamefont {Kahl}}]{bianchiJCP12}%
  \BibitemOpen
  \bibfield  {author} {\bibinfo {author} {\bibfnamefont {Emanuela}\
  \bibnamefont {Bianchi}}, \bibinfo {author} {\bibfnamefont {G\"{u}nther}\
  \bibnamefont {Doppelbauer}}, \bibinfo {author} {\bibfnamefont {Laura}\
  \bibnamefont {Filion}}, \bibinfo {author} {\bibfnamefont {Marjolein}\
  \bibnamefont {Dijkstra}}, \ and\ \bibinfo {author} {\bibfnamefont {Gerhard}\
  \bibnamefont {Kahl}},\ }\bibfield  {title} {\enquote {\bibinfo {title}
  {Predicting patchy particle crystals: Variable box shape simulations and
  evolutionary algorithms},}\ }\href {\doibase 10.1063/1.4722477} {\bibfield
  {journal} {\bibinfo  {journal} {J. Chem. Phys.}\ }\textbf {\bibinfo {volume}
  {136}},\ \bibinfo {eid} {214102} (\bibinfo {year} {2012})}\BibitemShut
  {NoStop}%
\bibitem [{\citenamefont {Phillips}\ and\ \citenamefont
  {Voth}(2013)}]{GregVoth2013}%
  \BibitemOpen
  \bibfield  {author} {\bibinfo {author} {\bibfnamefont {Carolyn~L.}\
  \bibnamefont {Phillips}}\ and\ \bibinfo {author} {\bibfnamefont {Gregory~A.}\
  \bibnamefont {Voth}},\ }\bibfield  {title} {\enquote {\bibinfo {title}
  {Discovering crystals using shape matching and machine learning},}\ }\href
  {\doibase 10.1039/C3SM51449H} {\bibfield  {journal} {\bibinfo  {journal}
  {Soft Matter}\ }\textbf {\bibinfo {volume} {9}},\ \bibinfo {pages}
  {8552--8568} (\bibinfo {year} {2013})}\BibitemShut {NoStop}%
\bibitem [{\citenamefont {Ashcroft}\ and\ \citenamefont
  {Mermin}(1976)}]{ashcroft1976solid}%
  \BibitemOpen
  \bibfield  {author} {\bibinfo {author} {\bibfnamefont {N.W.}\ \bibnamefont
  {Ashcroft}}\ and\ \bibinfo {author} {\bibfnamefont {N.D.}\ \bibnamefont
  {Mermin}},\ }\href@noop {} {\emph {\bibinfo {title} {{Solid State
  Physics}}}}\ (\bibinfo  {publisher} {Saunders College},\ \bibinfo {address}
  {Philadelphia},\ \bibinfo {year} {1976})\BibitemShut {NoStop}%
\bibitem [{\citenamefont {Prestipino}\ \emph {et~al.}(2009)\citenamefont
  {Prestipino}, \citenamefont {Saija},\ and\ \citenamefont
  {Malescio}}]{zerotempsanti}%
  \BibitemOpen
  \bibfield  {author} {\bibinfo {author} {\bibfnamefont {Santi}\ \bibnamefont
  {Prestipino}}, \bibinfo {author} {\bibfnamefont {Franz}\ \bibnamefont
  {Saija}}, \ and\ \bibinfo {author} {\bibfnamefont {Gianpietro}\ \bibnamefont
  {Malescio}},\ }\bibfield  {title} {\enquote {\bibinfo {title} {The
  zero-temperature phase diagram of soft-repulsive particle fluids},}\ }\href
  {\doibase 10.1039/B903931G} {\bibfield  {journal} {\bibinfo  {journal} {Soft
  Matter}\ }\textbf {\bibinfo {volume} {5}},\ \bibinfo {pages} {2795--2803}
  (\bibinfo {year} {2009})}\BibitemShut {NoStop}%
\bibitem [{Note3()}]{Note3}%
  \BibitemOpen
  \bibinfo {note} {Using the approach outlined in the text, the competitive
  pools determined for use in optimizations targeting the honeycomb lattice
  consisted of triangular, square, kagome, snub-square, elongated triangular,
  rectangular (\(b/a=1.49\)), rectangular (\(b/a=1.54\)), rectangular
  (\(b/a=1.56\)), rectangular (\(b/a=1.7\)), and snub-hexagonal lattices. For
  the square lattice, the final pool comprised triangular, oblique
  (\(b/a=1.514\), \(\theta \)=1.234), kagome, honeycomb, elongated triangular,
  snub-hexagonal, and snub-square lattices. For diamond, the pool~\cite
  {SMJainTMT2013} consisted of FCC, WUR, SH \((c/a=1.5)\), \(\beta \)Sn
  \((c/a=1.39)\), \(\beta \)Sn \((c/a=1.25)\), A7 \((b/a=3.79, u=0.1385)\), and
  A20 \((b/a=1.728, c/a=0.626,y=0.167)\) lattices. For simple cubic, the
  pool~\cite {SMJainTMT2013} comprised FCC, BCC, DIA, SH \((c/a=1)\), SH
  \((c/a=1.08)\), SH \((c/a=1.172)\), A20 \((b/a=1.72, c/a= 0.66, y= 0.67)\),
  \(\beta \)Sn \((c/a=0.873)\), \(\beta \)Sn \((c/a=0.78)\), and \(\beta \)Sn
  \((c/a=1.75)\) lattices. Here, \(b/a\) and \(c/a\) denote the aspect ratio of
  the sides of the unit cell, and \(\theta \) is the angle between the two
  sides. The other symbols \(u\) and \(y\), we adopt here, are the same as
  those used in a previous study~\cite {zerotempsanti}.}\BibitemShut {Stop}%
\bibitem [{\citenamefont {Jain}\ \emph
  {et~al.}(2013{\natexlab{b}})\citenamefont {Jain}, \citenamefont {Errington},\
  and\ \citenamefont {Truskett}}]{jainJCP}%
  \BibitemOpen
  \bibfield  {author} {\bibinfo {author} {\bibfnamefont {Avni}\ \bibnamefont
  {Jain}}, \bibinfo {author} {\bibfnamefont {Jeffrey~R.}\ \bibnamefont
  {Errington}}, \ and\ \bibinfo {author} {\bibfnamefont {Thomas~M.}\
  \bibnamefont {Truskett}},\ }\bibfield  {title} {\enquote {\bibinfo {title}
  {Communication: Phase behavior of materials with isotropic interactions
  designed by inverse strategies to favor diamond and simple cubic lattice
  ground states},}\ }\href {\doibase 10.1063/1.4825173} {\bibfield  {journal}
  {\bibinfo  {journal} {J. Chem. Phys.}\ }\textbf {\bibinfo {volume} {139}},\
  \bibinfo {eid} {141102} (\bibinfo {year} {2013}{\natexlab{b}})}\BibitemShut
  {NoStop}%
\bibitem [{Note4()}]{Note4}%
  \BibitemOpen
  \bibinfo {note} {Optimal parameters for the honeycomb-forming interaction
  \(\varphi _{\protect \text {hc}}\) are \(A=0.326914\), \(n=3.63306\),
  \(\lambda _{1}=0.286436\), \(k_{1}=3.6569\), \(\delta _{1}=1.26977\),
  \(\lambda _{2}=1.03258\), \(k_{2}=3.03683\), \(\delta _{2}=1.1557\),
  \(P=-0.175071\), \(Q=0.800825\), \(R=-0.937142\), and \(x_{\protect \text
  {cut}}=2.03291\). For the square-forming potential \(\varphi _{\protect \text
  {squ}}\), they are \(A=0.0946889\), \(n=3.53953\), \(\lambda _{1}=0.32989\),
  \(k_{1}=1.89197\), \(\delta _{1}=1.99003\), \(\lambda _{2}=0.062012\),
  \(k_{2}=5.89983\), \(\delta _{2}=1.0809\), \(P=0.433908\), \(Q=-2.4391\),
  \(R=3.46552\), and \(x_{\protect \text {cut}}=2.27813\).}\BibitemShut {Stop}%
\bibitem [{Note5()}]{Note5}%
  \BibitemOpen
  \bibinfo {note} {Also see Table S2 in Supplemental Information which
  tabulates the stable ground-state lattices with their corresponding density
  ranges and lattice parameters}\BibitemShut {NoStop}%
\bibitem [{\citenamefont {Marcotte}\ \emph
  {et~al.}(2011{\natexlab{b}})\citenamefont {Marcotte}, \citenamefont
  {Stillinger},\ and\ \citenamefont {Torquato}}]{2DMonotonicSM}%
  \BibitemOpen
  \bibfield  {author} {\bibinfo {author} {\bibfnamefont {E.}~\bibnamefont
  {Marcotte}}, \bibinfo {author} {\bibfnamefont {F.~H.}\ \bibnamefont
  {Stillinger}}, \ and\ \bibinfo {author} {\bibfnamefont {S.}~\bibnamefont
  {Torquato}},\ }\bibfield  {title} {\enquote {\bibinfo {title} {Optimized
  monotonic convex pair potentials stabilize low-coordinated crystals},}\
  }\href {\doibase 10.1039/C0SM01205J} {\bibfield  {journal} {\bibinfo
  {journal} {Soft Matter}\ }\textbf {\bibinfo {volume} {7}},\ \bibinfo {pages}
  {2332--2335} (\bibinfo {year} {2011}{\natexlab{b}})}\BibitemShut {NoStop}%
\bibitem [{Note6()}]{Note6}%
  \BibitemOpen
  \bibinfo {note} {Ground states and finite-temperature phase boundaries for
  \(\varphi _{\protect \text {dia}}\) and \(\varphi _{\protect \text {sc}}\)
  have been determined previously and are presented in detail elsewhere~\cite
  {SMJainTMT2013,jainJCP}.}\BibitemShut {Stop}%
\bibitem [{\citenamefont {van Meel}\ \emph
  {et~al.}(2009{\natexlab{a}})\citenamefont {van Meel}, \citenamefont
  {Frenkel},\ and\ \citenamefont {Charbonneau}}]{PRE2009CharbFrenk1}%
  \BibitemOpen
  \bibfield  {author} {\bibinfo {author} {\bibfnamefont {J.~A.}\ \bibnamefont
  {van Meel}}, \bibinfo {author} {\bibfnamefont {D.}~\bibnamefont {Frenkel}}, \
  and\ \bibinfo {author} {\bibfnamefont {P.}~\bibnamefont {Charbonneau}},\
  }\bibfield  {title} {\enquote {\bibinfo {title} {Geometrical frustration: A
  study of four-dimensional hard spheres},}\ }\href {\doibase
  10.1103/PhysRevE.79.030201} {\bibfield  {journal} {\bibinfo  {journal} {Phys.
  Rev. E}\ }\textbf {\bibinfo {volume} {79}},\ \bibinfo {pages} {030201}
  (\bibinfo {year} {2009}{\natexlab{a}})}\BibitemShut {NoStop}%
\bibitem [{\citenamefont {van Meel}\ \emph
  {et~al.}(2009{\natexlab{b}})\citenamefont {van Meel}, \citenamefont
  {Charbonneau}, \citenamefont {Fortini},\ and\ \citenamefont
  {Charbonneau}}]{PRE2009CharbFrenk2}%
  \BibitemOpen
  \bibfield  {author} {\bibinfo {author} {\bibfnamefont {J.~A.}\ \bibnamefont
  {van Meel}}, \bibinfo {author} {\bibfnamefont {B.}~\bibnamefont
  {Charbonneau}}, \bibinfo {author} {\bibfnamefont {A.}~\bibnamefont
  {Fortini}}, \ and\ \bibinfo {author} {\bibfnamefont {P.}~\bibnamefont
  {Charbonneau}},\ }\bibfield  {title} {\enquote {\bibinfo {title} {Hard-sphere
  crystallization gets rarer with increasing dimension},}\ }\href {\doibase
  10.1103/PhysRevE.80.061110} {\bibfield  {journal} {\bibinfo  {journal} {Phys.
  Rev. E}\ }\textbf {\bibinfo {volume} {80}},\ \bibinfo {pages} {061110}
  (\bibinfo {year} {2009}{\natexlab{b}})}\BibitemShut {NoStop}%
\end{thebibliography}
%

\clearpage
\newpage

\widetext
\begin{center}
\textbf{\large Supplemental Material: Dimensionality and design of isotropic interactions that stabilize honeycomb, square, simple cubic, and diamond lattices}
\end{center}

\setcounter{table}{0}
\setcounter{figure}{0}
\makeatletter
\renewcommand{\thetable}{S\arabic{table}}
\renewcommand{\thefigure}{S\arabic{figure}}

\begin{center}
\begin{table}[h]
\vspace{-2pt}
\caption{\label{tab:MCsimdetails} \small{\textbf{Parameters of the isochoric quenching Monte Carlo simulations.}\\ Notation (symbols) are as presented in the text. Errors in energy magnitudes are \(<O(-4)\).}}
 \begin{tabular}{cccccccccc}
 \hline
 \(\varphi_{\text{target}}\) & Target-structure & Simulation box & N & \(\rho\) & \(T_{\text{liq}}\) & \(T_{\text{xtal}}\) & \(E_{\text{conf}}\) & \(E_{\text{perf}}\) & \(\Delta E/E_{\text{perf}}\%\) \\
 \hline
 \(\vHc\)  & honeycomb    & 20 \(\times\) 20       & 800   & 1.23 & 0.1 & 0.02 & 3.149555 & 3.147587 & 0.062\\
 \(\vdia\) & honeycomb    & 20 \(\times\) 20       & 800   & 1.32 & 0.1 & 0.03 & 3.475706 & 3.473016 & 0.077\\
 \(\vSqu\) & square       & 20 \(\times\) 20       & 400   & 1.35 & 0.1 & 0.02 & 4.046236 & 4.045320 & 0.023\\ 
 \(\vSc\)  & square       & 20 \(\times\) 20       & 400   & 1.23 & 0.1 & 0.03 & 3.396865 & 3.396758 & 0.003\\
 \(\vHc\)  & diamond      & 8 \(\times\) 8 \(\times\) 8 & 1024 & 1.23 & 0.1 & 0.025 & 5.119002 & 5.114325 & 0.091 \\
 \(\vSqu\) & simple cubic & 8 \(\times\) 8 \(\times\) 8 & 512  & 1.77 & 0.1 & 0.01 & 10.225177 & 10.22514 & 0.004\\
 \hline
\end{tabular}{}
\vspace{-12pt}
\end{table}
\end{center}

\noindent
\emph{Isochoric quenching by Monte Carlo simulations}: We first completed a series of canonical Monte Carlo (MC) simulations to estimate the freezing behavior of each model by allowing a perfect lattice interacting with \(\varphi_{\text{target}}\), consisting of N particles at density \(\rho\), to relax at several temperatures separated by \(\Delta T=0.005\). We found that--for all interaction models--at \(T_{\text{liq}}=0.1\), the equilibrium structure was the fluid state. We instantaneously quenched these equilibrated disordered configurations to a much lower temperature value \(T_{\text{xtal}}\), and allowed the system to evolve for \(10^5\) MC steps. While some configurations instantaneously assembled to the expected crystal structure, there were also cases where configurations assembled into multiple high-energy defective structures before relaxing toward the final equilibrium structure. The pair distribution functions were averaged over approximately 4000 MC steps after assembly to the final equilibrium structure was achieved. We also simulated the expected crystal (per the ground-state phase diagram) at the same temperature \(T_{\text{xtal}}\) and density \(\rho\) to provide a comparison of the configurational energy of the quenched configuration \(E_{\text{conf}}\) versus the energy of the perfectly equilibrated target lattice \(E_{\text{perf}}\). 
We performed simulations with larger number of particles (upto 4050 particles for honeycomb crystal, 2025 for square crystal, 2500 for diamond crystal and 1300 for simple cubic crystal) for all lattices and also used cuboid box shapes, and found no significant differences in the crystallization behavior and free energies. For \(\vdia\), the fluid did not assemble into a perfect diamond crystal within the simulation time. We allowed the system to evolve for \(7 \times 10^5\) MC steps, and no change was seen in the lowest-energy configuration during the final cycle of \(10^5\) MC steps. However, as can be seen in the table, the energy difference between the perfect diamond crystal and the crystal formed on quenching is about 0.09\(\%\), and from Fig. 4 in main text, there are only very subtle discrepancies between the pair distribution function of the quenched configuration and the perfect diamond lattice.

\vspace{5pt}
\begin{center}  
\begin{longtable}{lll}
\caption{\label{tab:2DgsPD}\small{\textbf{2D ground states for optimized potentials \(\vHc\), \(\vSqu\), \(\vdia\) and \(\vSc\).}\\ Roman numerals denote different structures of the same lattice type.}} \\
  \hline
  Lattice & Stability range & Lattice parameters\\
 \hline
 \endfirsthead
 Lattice & Stability range & Lattice parameters\\
 \hline
 \multicolumn{3}{l}{\emph{Diamond forming potential}, \(\vdia\)} \\
   \hline
 \hline
 \endhead
    \multicolumn{3}{l}{\emph{Honeycomb forming potential}, \(\vHc\)} \\
   \hline
     \noalign{\smallskip}
   Triangular & \([0.44,0.77]\)\\
   Oblique-I & \([0.89:0.99]\) & \(b/a:[1.487:1.55]\)\\
   & & \(\theta:[1.227:1.24]\)\\ 
   Oblique-II & \([1.0:1.03]\) & \(b/a:1.71,\,\theta:1.05\)\\ 
   Rectangular & \([1.03:1.05]\) & \(b/a:1.49\) \\
   Honeycomb & \([1.11:1.37]\) \\
   Triangular & \([1.54:2.05]\) \\
   \noalign{\smallskip}
   \hline
   \multicolumn{3}{l}{\emph{Square forming potential}, \(\vSqu\)} \\
   \hline
   \noalign{\smallskip}
   Square & \([0.5:0.72]\) \\
   Triangular & \([0.75:0.95]\) \\
   Rectangular & \([0.95:1.11]\) & \(b/a:[1.43:1.45]\) \\ 
   Square & \([1.16:1.55]\) \\
   Elongated Triangular & \([1.58:1.70]\) \\
   Triangular & \([1.73:2.15]\) \\
   \noalign{\smallskip}
   \hline
   \multicolumn{3}{l}{\emph{Diamond forming potential}, \(\vdia\)} \\
   \hline
   \noalign{\smallskip}
   Triangular & \([0.46:0.83]\) \\
   Rectangular & \([0.97:1.12]\) & \(b/a:\,[1.446:1.526]\) \\
   Honeycomb & \([1.19:1.46]\) \\
   Kagome & \([1.55:1.58]\) \\
   Triangular & \([1.70:2.09]\) \\
    \noalign{\smallskip}
   \hline
   \multicolumn{3}{l}{\emph{Simple cubic forming potential}, \(\vSc\)} \\
   \hline
   \noalign{\smallskip}
   Triangular & \([0.5:1.0]\) \\
   Square & \([1.07:1.38]\) \\
   Triangular & \([1.45:1.8]\) \\
   \hline
 \end{longtable}
\end{center}

\vspace{5pt}
\begin{figure*}[h]
\centering
\includegraphics[scale=0.9]{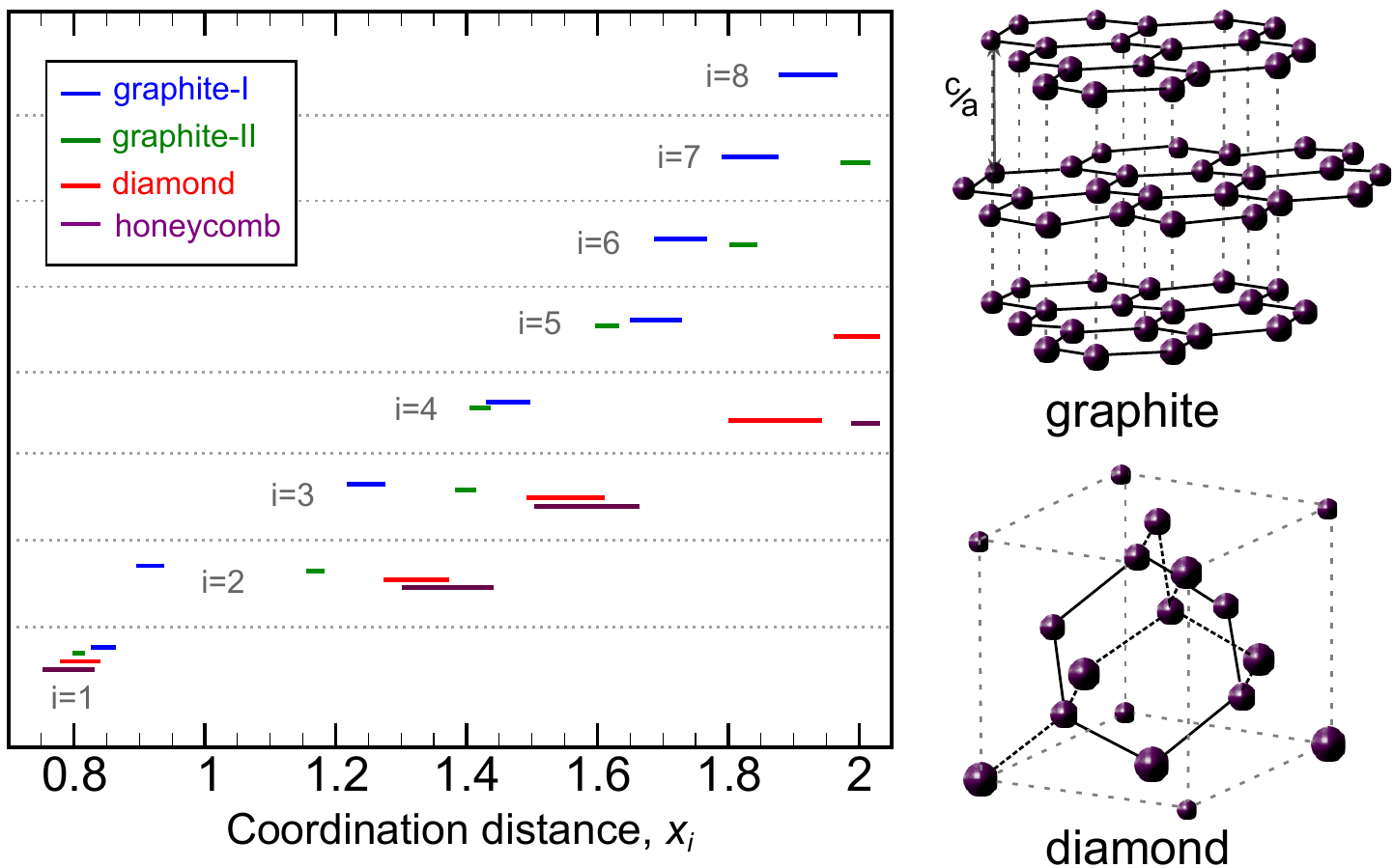}
\caption{\small{Interparticle separations corresponding to the \(i^{th}\) coordination shells (\(i\)=1, 2, 3...) of honeycomb, diamond, graphite-I, and graphite-II lattices within the interaction range for the potential \(\vHc\). Coordination-shell distances for honeycomb  \(\rho=[1.11,1.37]\) and diamond \(\rho=[1.09,1.38]\) correspond to their stability ranges for \(\vHc\) (see Fig.~3). For graphite-I (c/a=1.27) \(\rho=[1.1,1.26]\) and graphite-II (c/a=1.67) \(\rho=[0.98,1.05]\) lattices, the distances correspond to the density range at which each lattice is mechanically stable (i.e. the lowest phonon frequency has a positive value). We also highlight the honeycomb motif in the graphite and the diamond structures.}}
\label{fig:SuppCoordshells}
\end{figure*}

\noindent In Fig. S1, we plot the interparticle separations corresponding to the coordination shells of honeycomb, diamond and graphite structures within the interaction range for \(\vHc\). Graphite structures with axial ratios (c/a) in the ranges \([1.64,1.69]\) and \([1.25,1.27]\) are found to be optimal and mechanically stable in the density range of interest. However, on comparison with a larger pool of structures, A7-II and diamond (see Table I) have lower molar enthalpy and are chemically stable. We clearly see that there is no coordination-shell overlap beyond the nearest neighbour distances for all the four lattices, and hence, \(\vHc\) is not able to stabilize any of the graphite structures.

\end{document}